# ESTIMATING FUNCTIONS OF PROBABILITY DISTRIBUTIONS FROM A FINITE SET OF SAMPLES

## Part II: Bayes Estimators for Mutual Information, Chi-Squared, Covariance, and other Statistics.

by

David R. Wolf[1] and David H. Wolpert[2]

1 - Image Analysis Section, DX Division, DX-13, MS P940, LANL, Los Alamos, NM 87545. (wolf@lanl.gov)

2 - The Santa Fe Institute, 1660 Old Pecos Trail, Suite A, Santa Fe, NM 87501. (dhw@santafe.edu)

Abstract: This paper is the second in a series of two on the problem of estimating a function of a probability distribution from a finite set of samples of that distribution. In the first paper[1], the Bayes estimator for a function of a probability distribution was introduced, the optimal properties of the Bayes estimator were discussed, and the Bayes and frequency-counts estimators for the Shannon entropy were derived and graphically contrasted. In the current paper the analysis of the first paper is extended by the derivation of Bayes estimators for several other functions of interest in statistics and information theory. These functions are (powers of) the mutual information, chi-squared for tests of independence, variance, covariance, and average. Finding Bayes estimators for several of these functions requires extensions to the analytical techniques developed in the first paper, and these extensions form the main body of this paper. This paper extends the analysis in other ways as well, for example by enlarging the class of potential priors beyond the uniform prior assumed in the first paper. In particular, the use of the entropic and Dirichlet priors is considered.

## 1. BACKGROUND

Consider a system with m possible states and an associated m-vector of probabilities of those states, $\mathbf{p} = (p_i)$, $1 \le i \le m$, $(\Sigma_{i=1}^{m} p_i = 1)$. The system is repeatedly and independently sampled according to the distribution $\mathbf{p}$. Let the total number of samples be $N$ and denote the associated vector of counts of states by $\mathbf{n} = (n_i)$, $1 \le i \le m$, $(\Sigma_{i=1}^{m} n_i = N)$. By definition, $\mathbf{n}$ is multinomially distributed. In some cases in this paper the states will be indexed by two integers. For these cases $\mathbf{p}$ and $\mathbf{n}$ are matrices.

In many cases what we are interested in is not $\mathbf{p}$, but rather some function of $\mathbf{p}$, $F(\mathbf{p})$. The problem at hand is to estimate such a function $F(\mathbf{p})$ from the data $\mathbf{n}$. More precisely, the problem is to investigate the posterior density of $F(\mathbf{p})$, i.e. the probability density function (pdf)

$$P(F(\mathbf{p}) = f \mid \mathbf{n}) \equiv \int d\mathbf{p}\ \delta(F(\mathbf{p}) - f)\ P(\mathbf{p} \mid \mathbf{n}). \tag{1}$$

Usually it is difficult to analytically compute $P(F(\mathbf{p}) = f \mid \mathbf{n})$. Accordingly, here we instead compute posterior moments of $F(\mathbf{p})$, i.e., the moments of $F(\mathbf{p})$ according to the probability density in Eqn. (1).

These posterior moments do more than simply give us a characterization of $P(F(\mathbf{p}) = f \mid \mathbf{n})$. For example, in [1] we show that the estimator $G(\mathbf{n})$ that minimizes the mean-squared error from the true $F(\mathbf{p})$ is given by the first such posterior moment, the posterior average:

$$G(\mathbf{n}) = E[F(\mathbf{p}) \mid \mathbf{n}] \equiv \int d\mathbf{p}\ F(\mathbf{p})\ P(\mathbf{p} \mid \mathbf{n}) = \int df\ f\ P(f \mid \mathbf{n}). \tag{2}$$

In other words, this $G(\mathbf{n})$ minimizes

$$\int d\mathbf{p}\ P(\mathbf{p})\ \Sigma_{\mathbf{n}} P(\mathbf{n} \mid \mathbf{p}) \times (G(\mathbf{n}) - F(\mathbf{p}))^2, \tag{3}$$

where $P(\mathbf{p})$ is the so-called “prior” pdf of $\mathbf{p}$. In this sense, this $G(\mathbf{n})$ is the optimal estimator for $F(\mathbf{p})$. The estimator $E[F(\mathbf{p}) \mid \mathbf{n}]$ is known as the Bayes estimator for $F(\mathbf{p})$ with prior $P(\mathbf{p})$.

The second posterior moment is also useful. As was mentioned in [1], by using Chebyshev’s inequality, the second posterior moment can be used to bound the probability that $F(\mathbf{p})$ deviates substantially from the Bayes estimator for $F(\mathbf{p})$.

We now introduce the functions $F(\mathbf{p})$ for which we derive Bayes estimators in this paper. The functions $F(\mathbf{p})$ considered here are (the first two) powers of the mutual information, chi-squared for tests of independence, covariance, variance, and average. (In some cases the methods of this paper will allow consideration of all powers of these functions.) This choice of functions is not meant to be exhaustive. Rather it is meant to both exemplify some of the mathematical techniques involved in calculating Bayes estimators of function $F(\mathbf{p})$, and to provide some useful results. The techniques of this paper should be applicable to many other functions of interest as well.

The mutual information is defined in terms of a matrix $\mathbf{p}$ by

$$M(\mathbf{p}) = S((p_{i.})) + S((p_{.j})) - S(\mathbf{p}) \tag{4}$$

Here $(p_{i.})$ and $(p_{.j})$ are the vectors of column and row sums of $\mathbf{p} = (p_{ij})$ respectively, i.e. $p_{i.} \equiv \Sigma_j p_{ij}$ and similarly for $(p_{.j})$. $S(\mathbf{p})$ is the usual Shannon entropy: $S(\mathbf{p}) = -\Sigma_{ij} p_{ij} \log(p_{ij})$, while $S((p_{i.})) = -\Sigma_i p_{i.} \log(p_{i.})$, and similarly for $S((p_{.j}))$. Mutual information is a measure of the amount of information shared between two symbol streams (symbolic dynamical systems) with joint probability $p_{ij}$ [2]. It may also be seen as a measure of the correlation between two symbolic systems with joint probability $p_{ij}$ [3]. The mutual information function has applications in areas such as communication theory [2] (e.g. the measurement of channel capacity), pattern recognition [4], and natural languages [5], to name but a few.

The chi-squared statistic for independence is also given by a function of a matrix $\mathbf{p}$,

$$\chi^2(\mathbf{p}) \equiv \Sigma_{ij}((p_{ij} - p_{i.}p_{.j})^2 / (p_{i.}p_{.j})), \tag{5}$$

Chi-squared is commonly used in statistical tests of independence [6], where it appears in a form with the maximum-likelihood estimator of $\mathbf{p}$ substituted for $\mathbf{p}$ in Eq. (5). The form in Eq. (5) is proportional to the asymptotic (large data set) statistic used in these tests, and it is easily shown to be a first order approximation to the mutual information under certain conditions [7-8].

The covariance function of a matrix p is given by

$$\mathrm{Cov}_{xy}(\mathbf{p}) \equiv \Sigma_{ij} p_{ij} (X_i - \mu_x) (Y_j - \mu_y) , \tag{6}$$

where each of the m possible states is associated with some ordered pair $(X_i, Y_j)$ of numbers (there are m index pairs $(i, j)$ altogether). The $ij$'th state occurs with probability $p_{ij}$. The means are $\mu_x$ and $\mu_y$; $\mu_x \equiv \Sigma_i p_{i\cdot} X_i$ and similarly for $\mu_y$ [6].

The variance function of a vector **p** is given by

$$\mathrm{Var}(\mathbf{p}) \equiv \Sigma_i p_i (X_i - \mu_x)^2, \tag{7}$$

where the $i$'th state is associated with the number $X_i$ and occurs with probability $p_i$.

Finally, the average is a function of a vector **p** given by

$$\mathrm{Avg}_x(\mathbf{p}) \equiv \Sigma_i p_i X_i . \tag{8}$$

In Sec. 2 we derive Bayes estimators with uniform prior for these functions of probability distributions. The notation used in these results is summarized in subsections 2a and 2b. The results themselves, the Bayes estimators for the various functions being considered, are summarized in subsections 2d and 2g. In Sec. 3 we show how these Bayes estimator results along with those of [1], all derived under the assumption of a uniform prior, are modified when various different priors are assumed. In particular, we discuss the entropic prior $P(\mathbf{p}) \propto e^{\alpha S(\mathbf{p})}$ and a broad class of priors which includes the Dirichlet prior, $P(\mathbf{p}) \propto \Pi_{i=1}^{m} p_i^r$.

Throughout this paper it is assumed that the reader is at least passingly familiar with the analysis in [1], the first paper in this two-paper series.

## 2. CALCULATIONS FOR THE BAYES ESTIMATORS

In this section we present the calculations needed to derive the first two moments of the functions discussed in Sec. 1. The subsections are organized as follows. In Sec. 2a we discuss the form of the integrations to be done. Sec. 2b contains a presentation of notation motivated by considering the special case where $F(\mathbf{p})$ is the mutual information raised to some power. Section 2c contains intermediate results, Thms. 9-15. In Sec. 2d we use the results of Sec. 2c to derive the Bayes estimators for the first two powers of most of the functions described in Sec. 1. The results appear as Thms. 16-21. Sections 2e-g parallel Secs. 2b-d (notation, intermediate results (Thms. 22-25), and Bayes estimators (Thms. 26 and 27)), but for integrals more complicated than those considered in Secs. 2b-d.

The reader interested only in the results for the Bayes estimators for the $F(\mathbf{p})$ described in Sec. 1 should see Secs. 2d and 2g.

## 2a. THE FORM OF THE INTEGRALS.

Recall from Sec. 1 that the Bayes estimator for $F(\mathbf{p})$ is the posterior average $E[F(\mathbf{p})|\mathbf{n}] = \int d\mathbf{p}\ P(\mathbf{p}|\mathbf{n})\, F(\mathbf{p})$. Let $\Delta(\mathbf{p}) \equiv \delta(\Sigma_i p_i - 1)$ and $\Theta(\mathbf{p}) \equiv \Pi_i \theta(p_i)$, where $\theta(x) = 1$ for $x \geq 0$, 0 otherwise. Define $I[F(\mathbf{p}), \mathbf{n}]$ by

$$I[F(\mathbf{p}), \mathbf{n}] \equiv \int d\mathbf{p} F(\mathbf{p}) \Delta(\mathbf{p}) \Theta(\mathbf{p}) \Pi_{i=1}^{m} p_i^{n_i}. \qquad (9)$$

Using this notation, when the prior $P(\mathbf{p})$ is uniform, i.e. when $P(\mathbf{p}) \propto \Delta(\mathbf{p})\Theta(\mathbf{p})$, it is easily shown (see [1]) that

$$E[F(\mathbf{p})|\mathbf{n}] = I[F(\mathbf{p}), \mathbf{n}] / I[1, \mathbf{n}]. \qquad (10)$$

The result for $I[1, \mathbf{n}]$ appears in Thm. 3 of [1]. (In general, references to Thms. 1-8 are to the theorems so numbered in [1]; this paper, being a continuation of [1], starts numbering its theorems at 9.) Therefore, when the prior is uniform, finding the Bayes estimator $E[F(\mathbf{p})|\mathbf{n}]$ reduces to evaluating the integral $I[F(\mathbf{p}), \mathbf{n}]$. In the rest of this section, this integral is calculated for each of the $F(\mathbf{p})$ mentioned in Sec. 1.

## 2b. NOTATION.

To understand the types of calculations that must be performed, consider the case where $F(\mathbf{p}) = M^k(\mathbf{p})$, $M(\mathbf{p})$ being the mutual information. In this case the integral of interest is

$$I[M^k(\mathbf{p}), \mathbf{n}] = \int_{\Delta_\mathbf{P}} d\mathbf{p}\ \Pi_{i=1}^m p_i^{n_i} \times [-\Sigma_i p_{i\cdot} \log(p_{i\cdot}) - \Sigma_j p_{\cdot j} \log(p_{\cdot j}) + \Sigma_{ij} p_{ij} \log(p_{ij})]^k \quad (11)$$

(Here and in the rest of this paper we do not explicitly write the $\Delta(\mathbf{p})\Theta(\mathbf{p})$ factor in the integrand. Rather, it is indicated by subscripting the integrals with $\Delta_\mathbf{P}$.) The right-hand side of Eq. (11) expands to a sum of integrals of the form $I[\rho_1^{q_1} \log^{r_1}(\rho_1) \ldots \rho_k^{q_k} \log^{r_k}(\rho_k), \mathbf{n}]$, with $q_i = r_i$, and with each $\rho_i$ a sum of a subset of the $p_i$'s.

Since such sums of subsets of $p_i$'s will often arise, we introduce some special notation. Indicate a subset of indices of the $p_i$'s by $\sigma$, and the sum of the $p_i$'s with indices $i \in \sigma$ by $\rho \equiv \Sigma_{i \in \sigma} p_i$. If there are k such subsets, these will be represented by $\sigma_u$, $u = 1, \ldots, k$, and the corresponding subset sums will be given by $\rho_u \equiv \Sigma_{i \in \sigma_u} p_i$, $u = 1, \ldots, k$. In the case where $\sigma_u \cap \sigma_v = \varnothing$ for $u \neq v$, the subsets will be called *non-overlapping*. If on the other hand $\sigma_u \cap \sigma_v \neq \varnothing$ and yet $\sigma_u \neq \sigma_v$, the subsets will be called *overlapping*. Any expression involving non-overlapping (overlapping) subset sums will be called a non-overlapping (overlapping) term. *Pair-wise overlapping* will be used to indicate a term involving two subsets which are overlapping (as opposed to multiple such subsets).

Since we will often be dealing with overlapping subsets, the notation $\sigma_{uv}$ will be used for $\sigma_u \cap \sigma_v$. (No confusion arises since we never explicitly refer to $\sigma_u$ with u a double-digit number.) Similarly, we use the notation $\sigma_{u-uv} \equiv \sigma_u - \sigma_{uv}$ to refer to those indices in $\sigma_u$ but not in $\sigma_{uv}$, $\sigma_{u+v} \equiv \sigma_u \cup \sigma_v$ to refer to those indices in either $\sigma_u$ and/or in $\sigma_v$, and obvious extensions when more than two sets are being considered.

Note that in Eq. (11), for $k = 1$ the integral reduces to a sum of integrals of non-overlapping terms. For $k = 2$ some of the integrands are pair-wise overlapping terms. However for $k = 2$ no term occurs that involves more than two overlapping subsets.

We will use the following conventions for various other quantities which arise. For generic numbers appearing as exponents in convolution expressions the variable $\alpha$ will be used. Just as we previously defined $\rho$'s to refer to sums of those $p_i$ picked out by index subsets $\sigma$, we also need notation for the sums of $n_i$'s picked out by such $\sigma$'s. The variable $\beta$ will be used to indicate these sums, with the convention discussed above for subscripts on $\rho$ holding for $\beta$'s subscripts; $\beta_u \equiv \Sigma_{i \in \sigma_u} (n_i + 1)$. It will also sometimes be useful to use the notational convention $\nu_i \equiv n_i + 1$, $i = 1, \ldots, m$. In the case where **n** and **p** are matrices instead of vectors we use $\nu_{ij} \equiv n_{ij} + 1$. Similarly, to denote row or column sums of $n_{ij}$'s we use $\nu_{i\cdot} \equiv \Sigma_j \nu_{ij}$ and $\nu_{\cdot j} \equiv \Sigma_i \nu_{ij}$.

We will also have need of the following notation: $\beta \equiv \nu \equiv \Sigma_{i=1}^{m} \nu_i$; $\gamma_{\mathbf{n}} \equiv \Pi_{i=1}^{m} \Gamma(\nu_i)$; $\eta \equiv \Sigma_{u=1}^{k} \eta_u$ ($\eta_u$ is associated with subset $\sigma_u$, $u = 1, \ldots, k$); $\Phi^{(n)}(z) \equiv \Psi^{(n-1)}(z)$, where $\Psi^{(n)}(z)$ is given by $\Psi^{(n)}(z) = \partial_z^{n+1} \log(\Gamma(z))$ (see [9], Eq. 6.3.1); and the "delta-phi" function is given by $\Delta\Phi^{(n)}(z_1, z_2) \equiv \Phi^{(n)}(z_1) - \Phi^{(n)}(z_2)$.

Notation for the hypergeometric functions used here appears in App. A. In particular, the special use of subscripts on parentheses (e.g., $(a)_b$) is defined there, as are functions of the form ${}_pF_q$ and functions of the form ${}_{p_1, p_2, p_{12}}F_{q_1, q_2, q_{12}}$.

As it stands, the integrand in Eq. (11) is not "factorable" as defined in Sec. 4b of [1]. Having an integrand in factorable form is desirable because it allows us to apply the procedure of Sec. 4b of [1] to evaluate the integral. In what follows we utilize the T transform (see App. B.1) to convert the integrand of Eq. (11) (and similar integrands to appear) into factorable form. Once the integrand is factorable, we can perform the integration. After the integration we inverse T-transform to arrive at the final result for the integral of interest.

## 2c. NON-OVERLAP AND PAIR-WISE OVERLAP CONVOLUTION INTEGRALS.

In this subsection we derive a number of intermediate results. We begin this subsection with the fundamental convolution integrals needed to derive non-overlap and pair-wise overlap convolutions (Thms. 9-11). We then present specific integrals $I[\cdot,\cdot]$ of these two types (Thms. 12-15). Theorems 12 and 13a,b apply to non-overlap terms; Theorems 14a,b and 15a,b apply to pair-wise overlap terms. In Sec. 2d we use these intermediate results to evaluate Bayes estimators for several of the functions $F(\mathbf{p})$ being considered.

Define the Laplace convolution operator "⊗" acting on functions $f$ and $g$ by

$$(f \otimes g)(\tau) \equiv \int_0^\tau dx\ f(x)\, g(\tau - x)\,. \tag{12}$$

In the following, the convolution operator is (usually) implicitly assumed to refer to functions of the variable $p$. As described in [1], this convolution operator is fundamental to evaluating $I[\cdot,\cdot]$ integrals.

<u>Theorem 9.</u> If $\mathrm{Re}(\alpha_i) > 0,\ i = 1, 2$. Then

$$(9.1)\ \ (p^{\alpha_1 - 1} \otimes p^{\alpha_2 - 1})(\tau) = \frac{\Gamma(\alpha_1)\Gamma(\alpha_2)}{\Gamma(\alpha_1 + \alpha_2)} \times \tau^{\alpha_1 + \alpha_2 - 1},$$

$$(9.2)\ \ (p^{\alpha_1 - 1} e^{-pt} \otimes p^{\alpha_2 - 1} e^{-pt})(\tau) = \frac{\Gamma(\alpha_1)\Gamma(\alpha_2)}{\Gamma(\alpha_1 + \alpha_2)} \times \tau^{\alpha_1 + \alpha_2 - 1} \times e^{-\tau t}.$$

<u>Proof:</u> To Prove (9.1) note that by Thm. 2 (the Laplace convolution theorem) and the fact that $L[p^{\alpha - 1}](s) = \frac{\Gamma(\alpha)}{s^\alpha}$ for $\mathrm{Re}(\alpha) > 0$ we have

$$(p^{\alpha_1 - 1} \otimes p^{\alpha_2 - 1})(\tau) = L^{-1}[L[p^{\alpha_1 - 1}]\, L[p^{\alpha_2 - 1}]]\ (\tau)$$

$$= L^{-1}\left[\frac{\Gamma(\alpha_1)}{s^{\alpha_1}}\frac{\Gamma(\alpha_2)}{s^{\alpha_2}}\right](\tau) = \frac{\Gamma(\alpha_1)\Gamma(\alpha_2)}{\Gamma(\alpha_1+\alpha_2)} \times \tau^{\alpha_1+\alpha_2-1}.$$

To prove (9.2) note that $L\,[p^{\alpha-1}e^{-pt}](s) = \dfrac{\Gamma(\alpha)}{(s+t)^{\alpha}}$. The remainder of the proof parallels the proof of (9.1) above: Substitute $L\,[p^{\alpha-1}e^{-pt}]$ for $L\,[p^{\alpha}]$. QED.

If desired, Thm. 3 may be thought of as corollary to Thm. 9.1, by induction.

In order to discuss the next results succinctly, we need to use the confluent hypergeometric function, discussed in App. A.

Theorem 10. If $\mathrm{Re}\,(\alpha_1) > 0$, $\mathrm{Re}\,(\alpha_2) > 0$, and $\bar{\alpha} = \alpha_1 + \alpha_2$ then

(10.1)

$$(p^{\alpha_1-1}e^{-pt_1} \otimes p^{\alpha_2-1}e^{-p(t_1+t_2)})(\tau) = \frac{\Gamma(\alpha_1)\Gamma(\alpha_2)}{\Gamma(\bar{\alpha})} \times \tau^{\bar{\alpha}-1} \times e^{-\tau(t_1+t_2)}$$

$$\times\, {}_1F_1(\alpha_1;\ \bar{\alpha};\ t_2\tau)$$

(10.2)

$$(p^{\alpha_1-1} \otimes p^{\alpha_2-1}e^{-pt})(\tau) = \frac{\Gamma(\alpha_1)\Gamma(\alpha_2)}{\Gamma(\bar{\alpha})} \times \tau^{\bar{\alpha}-1} \times e^{-\tau t} \times {}_1F_1(\alpha_1;\ \bar{\alpha};\ t\tau)$$

Proof: To prove (10.1) write the convolution $(p^{\alpha_1-1}e^{-pt_1} \otimes p^{\alpha_2-1}e^{-p(t_1+t_2)})(\tau)$ in its integral form as $e^{-\tau(t_1+t_2)}\int_0^{\tau} e^{pt_2}\, p^{\alpha_1-1}\,(\tau-p)^{\alpha_2-1}dp$. Make the change of variables $\tau u = p$, which introduces the factor $\tau^{\bar{\alpha}-1}$. Substitute $\alpha = \alpha_1$, $\beta = \bar{\alpha}$, $z = t_2\tau$ and compare with the form in App. A to find the result. The factor $\dfrac{\Gamma(\beta)}{\Gamma(\alpha)\Gamma(\beta-\alpha)}$ in the hypergeometric function in App. A cancels the factor $\frac{\Gamma(\alpha_1)\Gamma(\alpha_2)}{\Gamma(\alpha)}$. To prove (10.2) substitute zero for $t_1$ in result (10.1).

QED.

Theorem 11 gives the primary equality needed to derive pair-wise overlap results. See App. A for the definition of the generalized hypergeometric function ${}_{1,1,0}F_{0,0,1}$.

<u>Theorem 11.</u> If $\mathrm{Re}(\alpha_1) > 0$, $\mathrm{Re}(\alpha_{12}) > 0$, $\mathrm{Re}(\alpha_2) > 0$ and $\alpha \equiv \alpha_1 + \alpha_{12} + \alpha_2$, then

$$(p^{\alpha_1 - 1}e^{-pt_1} \otimes p^{\alpha_{12} - 1}e^{-p(t_1 + t_2)} \otimes p^{\alpha_2 - 1}e^{-pt_2})(\tau) =$$

$$\frac{\Gamma(\alpha_1)\Gamma(\alpha_{12})\Gamma(\alpha_2)}{\Gamma(\alpha)} \times \tau^{\alpha - 1} \times e^{-\tau(t_1 + t_2)} \times {}_{1,1,0}F_{0,0,1}[\alpha_1, \alpha_2;\ \alpha;\ t_2\tau, t_1\tau]$$

<u>Proof</u>: The convolution $p^{\alpha_1 - 1}e^{-pt_1} \otimes p^{\alpha_{12} - 1}e^{-p(t_1 + t_2)}$ was done in Thm. 10.1. Expand ${}_1F_1$ in that result in its series representation (see App. A)

$${}_1F_1(\alpha_1;\ \alpha_1 + \alpha_{12};\ t_2\tau) = \Sigma_{k=0}^{\infty} \frac{(\alpha_1)_k}{(\alpha_1 + \alpha_{12})_k} \frac{(t_2\tau)^k}{\Gamma(k+1)}$$

and convolve it with $p^{\alpha_2 - 1}e^{-pt_2}$ term-by-term (apply Thm. 10.2) to find the desired result. (This is valid since the series is uniformly convergent on $[0, \tau]$.) QED.

Thms. 9-11 are now used to derive integrals $I[\cdot,\cdot]$ for some non-overlap and pair-wise overlap terms. Theorems 12 and 13 discuss the non-overlap case, while Thms. 14 and 15 discuss the pair-wise overlap case.

<u>Theorem 12.</u> If the subsets $\sigma_u$, defined for $u = 1, \ldots, k$, satisfy $\sigma_{uv} = \varnothing$ for all $u \neq v$, and if $\mathrm{Re}(\beta_u + \eta_u) > 0$ for all $u = 1, \ldots, k$, and if $\mathrm{Re}(\nu_i) > 0$ for all $i = 1, \ldots, m$ then

$$I[\rho_1^{\eta_1}\ldots\rho_k^{\eta_k}, \mathbf{n}] = \frac{\gamma_{\mathbf{n}}}{\Gamma(\beta+\eta)} \Pi_{u=1}^{k} \frac{\Gamma(\beta_u+\eta_u)}{\Gamma(\beta_u)}.$$

Proof: Assume $k = 1$ and $\mathrm{Re}(\eta_1) < 0$ to begin. Apply $T^{-1}T$ with respect to $\eta_1$ to the integral $I[\rho_1^{\eta_1}, \mathbf{n}]$ (see App. B.1 for the definition of the T transform) and evaluate the inner T transform. Noting $e^{-\rho_1 t} = \Pi_{i \in \sigma_1} e^{-p_i t}$ and that $T[z^{\eta}](t) = e^{-zt}$ for $\eta_1 < 0$, find

$$I[\rho_1^{\eta_1}, \mathbf{n}] = T^{-1}\left[ \int_{\Delta_{\mathbf{p}}} d\mathbf{p}\, (\Pi_{i \in \sigma_1} p_i^{n_i} e^{-p_i t}) \times (\Pi_{i \notin \sigma_1} p_i^{n_i}) \right].$$

(See App. C.1 for the justification of the interchange the integral over $\mathbf{p}$ and the T transform.) Now, write the transformed integral above as the convolution

$$I[\rho_1^{\eta_1}, \mathbf{n}] = T^{-1}((\otimes_{i \in \sigma_1} p_i^{n_i} e^{-p_i t}) \otimes (\otimes_{i \notin \sigma_1} p_i^{n_i}))(1).$$

Use Thm. 9.1 and induction to find (with $\bar{\beta} \equiv \beta - \beta_1$)

$$(\otimes_{i \notin \sigma_1} p_i^{n_i})(\tau) = \frac{\Pi_{i \notin \sigma_1} \Gamma(\nu_i)}{\Gamma(\bar{\beta})} \times \tau^{\bar{\beta}-1}.$$

Similarly, use Thm. 9.2 and induction to find

$$(\otimes_{i \in \sigma_1} p_i^{n_i} e^{-p_i t})(\tau) = \frac{\Pi_{i \in \sigma_1} \Gamma(\nu_i)}{\Gamma(\beta_1)} \times \tau^{\beta_1 - 1} \times e^{-\tau t}.$$

Substituting the last two expressions into $I[\rho_1^{\eta_1}, \mathbf{n}]$ yields

$$I[\rho_1^{\eta_1}, \mathbf{n}] = \frac{\gamma_{\mathbf{n}}}{\Gamma(\beta_1)\Gamma(\bar{\beta})} T^{-1}(\tau^{\beta_1 - 1} e^{-\tau t} \otimes \tau^{\bar{\beta}-1})(1).$$

The $T^{-1}$ transform may now be taken to find (see Apps. A, C)

$$I[\rho_1^{\eta_1}, \mathbf{n}] = \frac{\gamma_{\mathbf{n}}}{\Gamma(\beta_1)\Gamma(\bar{\beta})} (\tau^{\beta_1 + \eta_1 - 1} \otimes \tau^{\bar{\beta}-1})(1).$$

Now apply Thm. 9.1 in this expression to find for $\eta_1 < 0$

$$I\,[\rho_1^{\eta_1}, \mathbf{n}] \;=\; \frac{\gamma_{\mathbf{n}}\Gamma(\beta_1+\eta_1)}{\Gamma(\beta_1)\Gamma(\beta+\eta)}.$$

Refer to App. D for the continuation to $\mathrm{Re}\,(\eta_1) \geq 0$. Refer to App. E for the existence conditions. Now, for $k > 1$ apply the identity operator $T^{-1}T$ $k$ times (with respect to $\eta_1 \ldots \eta_k$ respectively) and evaluate only the T transforms initially. Since $\sigma_{uv} = \varnothing$ for $u \neq v$, the convolution form of the transformed $I\,[\rho_1^{\eta_1}, \mathbf{n}]$ becomes

$$I\,[\rho^{\eta_1}\ldots\rho^{\eta_k}, \mathbf{n}] \;=\; T_1^{-1}\ldots T_k^{-1}\big(\,(\otimes_{u=1}^{k}\,(\otimes_{i\,\in\,\sigma_u} p_i^{n_i} e^{-p_i t_u})\,)\otimes(\otimes_{i\,\notin\,\cup_{u=1}^{k}\sigma_u} p_i^{n_i})\,\big)(1),$$

while $\bar{\beta}$ becomes $\bar{\beta} \equiv \beta - \beta_{1+\ldots+k}$. Extend the application of Thm. 9.2 to the $k$ convolution products $\otimes_{i\,\in\,\sigma_u} p_i^{n_i} e^{-p_i t}$ for $u = 1, \ldots, k$. Do the substitutions and take the inverse transforms to find the result. QED.

Footnote 1 contains a derivation of an interesting identity based on an alternate form of the result in Thm. 12. Since observations being integer counts in practice, we'll be interested in non-negative integer $n_i$; in this regard, Thm. 12 is more general than we need. (However, below we will want to differentiate with respect to $\eta$'s, to get logarithms into the integrand. So a result only applicable to integer $\eta$'s would not suffice.)

Theorem 13 applies Thm. 12 to find non-overlap results needed specifically for the expression of Bayes estimators for the first two moments of the entropy, mutual information, and various other functions.

Theorem 13. If the subsets $\sigma_u$, defined for $u = 1, \ldots, k$, satisfy $\sigma_{uv} = \varnothing$ for all $u \neq v$, and if $\mathrm{Re}\,(\beta_u + \eta_u) > 0$ for all $u = 1, \ldots, k$, and if $\mathrm{Re}\,(\nu_i) > 0$ for all $i = 1, \ldots, m$ then the following hold.

(13.1) One logarithm of a subset sum.

$$I\,[\,(\rho_1^{\eta_1}\ldots\rho_k^{\eta_k})\times\log(\rho_u),\mathbf{n}]\;=\;\frac{\gamma_{\mathbf{n}}}{\Gamma(\beta+\eta)}\times\Pi_{w=1}^{k}\frac{\Gamma(\beta_w+\eta_w)}{\Gamma(\beta_w)}\times\Delta\Phi^{(1)}(\beta_u+\eta_u,\beta+\eta).$$

(13.2) Two logarithms of subset sums, $u \neq v$.

$$I\,[\,(\rho_1^{\eta_1}\ldots\rho_k^{\eta_k})\times\log(\rho_u)\times\log(\rho_v),\mathbf{n}]\;=\;\frac{\gamma_{\mathbf{n}}}{\Gamma(\beta+\eta)}\times\Pi_{w=1}^{k}\frac{\Gamma(\beta_w+\eta_w)}{\Gamma(\beta_w)}$$

$$\times\,\{\Delta\Phi^{(1)}(\beta_u+\eta_u,\beta+\eta)\Delta\Phi^{(1)}(\beta_v+\eta_v,\beta+\eta)-\Phi^{(2)}(\beta+\eta)\}\,.$$

(13.3) Squared logarithm of a subset sum.

$$I\,[\,(\rho_1^{\eta_1}\ldots\rho_k^{\eta_k})\times\log(\rho_u)^2,\mathbf{n}]\;=\;\frac{\gamma_{\mathbf{n}}}{\Gamma(\beta+\eta)}\times\Pi_{w=1}^{k}\frac{\Gamma(\beta_w+\eta_w)}{\Gamma(\beta_w)}$$

$$\times\,\{\Delta\Phi^{(1)}(\beta_u+\eta_u,\beta+\eta)^2+\Delta\Phi^{(2)}(\beta_u+\eta_u,\beta+\eta)\}\,.$$

Proof: The proof is done for result (13.1); results (13.2) and (13.3) follow in a similar manner. Differentiate both sides of the formula for $I\,[\rho^{\eta_1}\ldots\rho^{\eta_k},\mathbf{n}]$ given in Thm. 12 with respect to $\eta_u$ using the fact that $\partial_\eta\rho^\eta = \rho^\eta\log(\rho)$. (See App. C.2 for justification of the interchange the integral and derivative.) Doing this gives the desired result. QED.

Theorems 12 and 13 dealt with non-overlap sums. Theorems 14 and 15 below discuss pair-wise overlap sums. In Thm. 14a the non-contained overlap case is discussed. In Thm. 14b the contained overlap case is discussed. See App. A for the definition of the hypergeometric function ${}_{2,2,0}F_{0,0,1}$.

Theorem 14a. If $\sigma_1$ and $\sigma_2$ satisfy $\sigma_{12}\neq\varnothing$, $\sigma_1\neq\sigma_{12}\neq\sigma_2$, $\mathrm{Re}\,(\beta_1+\eta_1)>0$, $\mathrm{Re}\,(\beta_2+\eta_2)>0$, and $\mathrm{Re}\,(\nu_i)>0$, $i = 1, \ldots, m$, then

$$I\,[\rho_1^{\eta_1}\rho_2^{\eta_2},\mathbf{n}]\;=\;\frac{\gamma_{\mathbf{n}}}{\Gamma(\beta+\eta)}\times\frac{\Gamma(\beta_{1+2}+\eta)}{\Gamma(\beta_{1+2})}$$

$$\times\,{}_{2,2,0}F_{0,0,1}\,[\,(\beta_{1-12},-\eta_2),\,(\beta_{2-12},-\eta_1)\,;\beta_{1+2};1,1]\,.$$

Proof: To begin, assume that $\mathrm{Re}(\eta_i) < 0$, $i = 1, 2$ and that the $\eta_i$ are not integers. Apply $T_1^{-1}T_2^{-1}T_1T_2$ ($T_i$ is with respect to $\eta_i$, see App. B.1) to the integral $I[\rho_1^{\eta_1}\rho_2^{\eta_2}, \mathbf{n}]$. Evaluating the (non-inverse) T transforms yields the factorable form

$$I[\rho_1^{\eta_1}\rho_2^{\eta_2}, \mathbf{n}] = T_1^{-1}T_2^{-1}\int_{\Delta_{\mathbf{P}}} d\mathbf{p}\,(\Pi_{i\in(\sigma_{1-12})} p_i^{n_i} e^{-p_i t_1}) \times (\Pi_{i\in\sigma_{12}} p_i^{n_i} e^{-p_i(t_1+t_2)}) \times (\Pi_{i\in(\sigma_{2-12})} p_i^{n_i} e^{-p_i t_2}) \times (\Pi_{i\notin\sigma_{1+2}} p_i^{n_i}).$$

(See App. C.1 for justification of the interchange of the integral over $\mathbf{p}$ and the T transform.) Now, write the transformed integral as the convolution (see Thm. 1)

$$I[\rho_1^{\eta_1}\rho_2^{\eta_2}, \mathbf{n}] = T_1^{-1}T_2^{-1}((\otimes_{i\in\sigma_{1-12}} p_i^{n_i} e^{-p_i t_1}) \otimes (\otimes_{i\in\sigma_{12}} p_i^{n_i} e^{-p_i(t_1+t_2)}) \otimes (\otimes_{i\in\sigma_{2-12}} p_i^{n_i} e^{-p_i t_2}) \otimes (\otimes_{i\notin\sigma_{1+2}} p_i^{n_i}))(1).$$

Apply Thm. 9.1 and induction to find (where $\bar{\beta} \equiv \beta - \beta_{1+2}$)

$$(\otimes_{i\notin\sigma_{1+2}} p_i^{n_i})(\tau) = \frac{\Pi_{i\notin\sigma_{1+2}}\Gamma(\nu_i)}{\Gamma(\bar{\beta})} \times \tau^{\bar{\beta}-1}.$$

Similarly, use Thm. 9.2 and induction to find

$$((\otimes_{i\in\sigma_{1-12}} p_i^{n_i} e^{-p_i t_1}) \otimes (\otimes_{i\in\sigma_{12}} p_i^{n_i} e^{-p_i(t_1+t_2)}) \otimes (\otimes_{i\in\sigma_{2-12}} p_i^{n_i} e^{-p_i t_2}))(\tau) =$$
$$\frac{\Pi_{i\in\sigma_{1+2}}\Gamma(\nu_i)}{\Gamma(\beta_{1-12})\Gamma(\beta_{12})\Gamma(\beta_{2-12})}(p^{\beta_{1-12}-1}e^{-pt_1} \otimes p^{\beta_{12}-1}e^{-p(t_1+t_2)} \otimes p^{\beta_{2-12}-1}e^{-pt_2})(\tau).$$

Substitute the result for Thm. 11 into the triple convolution above, and substitute the last two expressions into the convolution form of the transformed integral to find

$$I[\rho_1^{\eta_1}\rho_2^{\eta_2}, \mathbf{n}] = T_1^{-1}T_2^{-2}\Big[\frac{\gamma_{\mathbf{n}}}{\Gamma(\beta_{1+2})\Gamma(\bar{\beta})} \times ((p^{\beta_{1+2}-1} \times e^{-p(t_1+t_2)} \times {}_{1,1,0}F_{0,0,1}[\beta_{1-12}, \beta_{2-12}; \beta_{1+2}; t_2 p, t_1 p]) \otimes p^{\bar{\beta}-1})\Big](1).$$

Now, take inverse T transforms and apply Thm. 9.1 to find the desired result. Refer to App. E to determine the conditions for the existence of the identity. Refer to App. D for the continuation of the result to $\mathrm{Re}\,(\eta_i) \geq 0$. Finally, for values of $\eta \geq 0$, refer to App. F. QED.

See footnote 2 for a derivation of two interesting identities resulting from alternate forms of this proof. When $\sigma_2 \subset \sigma_1$ the above result simplifies as in Thm. 14b below.

Theorem 14b. If $\sigma_1, \sigma_2$ satisfy $\sigma_{12} \neq \emptyset$, $\sigma_2 \subset \sigma_1$, $\mathrm{Re}\,(\beta_1 + \eta) > 0$, $\mathrm{Re}\,(\beta_{12} + \eta_2) > 0$, and $\mathrm{Re}\,(\nu_i) > 0, i = 1, \ldots, m$, then

$$\text{(14b.1)}\quad I\,[\rho_1^{\eta_1}\rho_2^{\eta_2}, \mathbf{n}] = \frac{\gamma_{\mathbf{n}}}{\Gamma(\beta+\eta)} \times \frac{\Gamma(\beta_1+\eta)}{\Gamma(\beta_1)} \times {}_2F_1\,[\,(\beta_{1-12}, -\eta_2)\,;\beta_1;1]\,,$$

$$\text{(14b.2)}\quad = \frac{\gamma_{\mathbf{n}}}{\Gamma(\beta_{12})} \times \frac{\Gamma(\beta_{12}+\eta_2)}{\Gamma(\beta_1+\eta_2)} \times \frac{\Gamma(\beta_1+\eta)}{\Gamma(\beta+\eta)}\,.$$

Proof: Similar to proof of Thm. 14a, but apply Thm. 10.1 instead of Thm. 11. The second form (14b.2) of the result is derived by applying Gauss's identity (see Footnote 1) to the first form of the result above. QED.

Theorems 15a and 15b build upon the results of Thms. 14a and 14b respectively and state results needed to express specific terms of the various Bayes estimators. Theorem 15a contains results for the non-contained overlap case. ("Non-contained" means that neither subset of indices is properly contained within the other.) Thm. 15b contains results for the contained overlap case. Since we are most directly interested in non-negative integer $\eta$'s and because simplification occurs at those $\eta$'s, Thm. 15a is stated only for non-negative integer $\eta$'s.

Theorem 15a. If $\eta_1 \geq 0$ and $\eta_2 \geq 0$ are integers and the conditions for Thm. 14a hold, then

$$\text{(15a.1)}\quad I\,[\rho_1^{\eta_1}\rho_2^{\eta_2}, \mathbf{n}] = C^{(0)}F^{(00)}\,,$$

(15a.2) $\quad I\,[\rho_1^{\eta_1}\log(\rho_1)\,\rho_2^{\eta_2},\mathbf{n}] = C^{(1)}F^{(00)} + C^{(0)}F^{(10)}$,

(15a.3) $\quad I\,[\rho_1^{\eta_1}\log(\rho_1)\,\rho_2^{\eta_2}\log(\rho_2),\mathbf{n}] =$

$$C^{(2)}F^{(00)} + C^{(1)}\,[F^{(10)} + F^{(01)}] + C^{(0)}F^{(11)},$$

(15a.4) $\quad I\,[\rho_1^{\eta_1}\log^2(\rho_1)\,\rho_2^{\eta_2},\mathbf{n}] = C^{(2)}F^{(00)} + 2C^{(1)}F^{(10)} + C^{(0)}F^{(20)}$,

where:

$$C^{(0)} \equiv \frac{\gamma_{\mathbf{n}}}{\Gamma(\beta_{1+2})} \times \frac{\Gamma(\beta_{1+2}+\eta)}{\Gamma(\beta+\eta)},\; C^{(1)} \equiv C^{(0)} \times \Delta\Phi^{(1)}(\beta_{1+2}+\eta,\beta+\eta),$$

$$C^{(2)} \equiv C^{(0)} \times \{\Delta\Phi^{(1)}(\beta_{1+2}+\eta,\beta+\eta)^2 + \Delta\Phi^{(2)}(\beta_{1+2}+\eta,\beta+\eta)\},$$

and

(a) $$F^{(00)} \equiv \eta_1!\eta_2!\Sigma_{i=0}^{\eta_2}\Sigma_{j=0}^{\eta_1}\frac{(\beta_{1-12})_i(\beta_{2-12})_j}{(\beta_{1+2})_{i+j}}\frac{1}{(\eta_1-j)!\,(\eta_2-i)!}\frac{(-1)^{i+j}}{i!j!},$$

(b) $$F^{(10)} \equiv \eta_1!\eta_2!\Sigma_{i=0}^{\eta_2}\Sigma_{j=0}^{\infty}\frac{(\beta_{1-12})_i(\beta_{2-12})_j}{(\beta_{1+2})_{i+j}}\frac{1}{(\eta_2-i)!}\frac{(-1)^{i}}{i!j!}Q_1(j,\eta_1),$$

with $Q_1$ given by

$$Q_1(j,\eta_1) \equiv (1-\theta(j-\eta_1-1))\frac{(-1)^j}{(\eta_1-j)!}\Sigma_{r=0}^{j-1}\frac{1}{\eta_1-r}$$

$$+\,\theta(j-\eta_1-1)\,(-1)^{\eta_1+1}\Gamma(j-\eta_1).$$

(c) $F^{(01)}$ is the same as (b) with $i \leftrightarrow j$ and $\eta_1 \leftrightarrow \eta_2$.

(d) $$F^{(11)} \equiv \eta_1!\eta_2!\Sigma_{i=0}^{\infty}\Sigma_{j=0}^{\infty}\frac{(\beta_{1-12})_i(\beta_{2-12})_j}{(\beta_{1+2})_{i+j}}\frac{1}{i!j!}Q_1(j,\eta_1)Q_1(i,\eta_2).$$

(e) $$F^{(20)} \equiv \eta_1!\eta_2!\Sigma_{i=0}^{\eta_2}\Sigma_{j=0}^{\infty}\frac{(\beta_{1-12})_i(\beta_{2-12})_j}{(\beta_{1+2})_{i+j}}\frac{1}{(\eta_2-i)!}\frac{(-1)^{i}}{i!j!}Q_2(j,\eta_1),$$

with $Q_2$ given by

$$Q_2(j,\eta_1) \equiv \{ (1-\theta(j-\eta_1-1)) \frac{(-1)^j}{(\eta_1-j)!} \Sigma_{r,s=0,r\neq s}^{j-1} \frac{1}{(\eta_1-r)(\eta_1-s)}$$

$$+\theta(j-\eta_1-1)(-1)^{\eta_1+1} 2\Gamma(j-\eta_1) \Sigma_{r=0,r\neq\eta_1}^{j-1} \frac{1}{\eta_1-r} \} .$$

<u>Proof:</u> The proof is done for (15a.2). The other cases have similar proofs. Differentiate both sides of the expression for $I[\rho_1^{\eta_1}\rho_2^{\eta_2}, \mathbf{n}]$ given in Thm. 14a with respect to $\eta_1$. After differentiating, the left-hand side is given by $I\left[\rho_1^{\eta_1}\log(\rho_1)\rho_2^{\eta_2}, \mathbf{n}\right]$. (The justification of the interchange of the integral and derivative is given in App. C.1.) Write the differentiated right-hand side as

$$\partial_{\eta_1}(C^{(0)} \times {}_{2,2,0}F_{0,0,1}[(\beta_{1-12}, -\eta_2), (\beta_{2-12}, -\eta_1); \beta_{1+2}; 1, 1]).$$

This expands to $\partial_{\eta_1}(C^{(0)})_{2,2,0}F_{0,0,1}[\ldots] + C^{(0)}\partial_{\eta_1}({}_{2,2,0}F_{0,0,1}[\ldots])$. The derivative of $C^{(0)}$ is given by $\partial_{\eta_1}C^{(0)} = C^{(0)} \times \Delta\Phi^{(1)}(\beta_{1+2}+\eta, \beta+\eta) = C^{(1)}$. The undifferentiated hypergeometric is evaluated at $\eta_1$ and $\eta_2$ using the results in App. F, cases 1 and 2. This evaluates to $F^{(00)}$ defined in (a) above. The derivative of the hypergeometric may be taken term-by-term (this is justified below). Use the results in App. F, cases 1 and 2, Eqs. (F.4) and (F.6), to evaluate this derivative at $\eta_1$ and $\eta_2$. Doing this gives the expression $F^{(10)}$ defined in (b). With these derivatives and evaluations, the result (15a.2) follows immediately. Now consider the validity of term-by-term differentiation of the hypergeometric. There exists a closed neighborhood N containing the integer $\eta_1$ with $\mathrm{Re}(\beta_1 + x) > 0 \ \forall x \in N$. The results of App. F show that any truncation (in $j$) of the series for ${}_{2,2,0}F_{0,0,1}[(\beta_{1-12}, -\eta_2), (\beta_{2-12}, -x); \beta_{1+2}; 1, 1]$ (see App. A) may be differentiated with respect to $x$ on $N$. The sequence of derivatives of the increasing order truncations converges uniformly on $N$. (To see this, note that $S_i(x) \equiv \Sigma_{j=\eta_1+1}^{\infty} \frac{\Gamma(\beta_{2-12}+j)}{\Gamma(\beta_{1+2}+i+j)} \frac{\Gamma(-x+j)}{j!}$ is convergent for each $i$, and $\mathrm{Re}(\beta_1 + x) > 0$. Now, note that $S_i(x)$ is a series of terms each monotonic on $N$ with the same monotonicity in $x$ holding for each term, and that the summation over $i$ in (b) is finite. These observations and the convergence just established demonstrate the claim of uniform convergence.) Finally, by Thm. 7.17 of [10], the sequence of derivatives of the increasing order truncations converges to the derivative of the limit of the series on $N$, justifying the term-by-term differentiation of the infinite series. QED.

See footnote 3 for some comments regarding alternate forms for the results given above in Thm. 15a.

Theorem 15b builds on Thm. 14b and states the results for the case in which there are two subset sums, with the indices of one subset completely contained in the other. Here, unlike in Thm. 15a, there is no hypergeometric function to consider, so the presentation of these results is much shorter. Further, unlike Thm. 15a, the expressions given are valid for all $\eta$'s in the range specified (not just at nonnegative integers as in Thm. 15a) because there are no poles in the expressions being considered at the integers and therefore no further simplification occurs at these points.

Theorem 15b. If the conditions for Thm. 14b hold, then

(15b.1) $I\left[\rho_1^{\eta_1}\log(\rho_1)\rho_2^{\eta_2}, \mathbf{n}\right] = C^{(00)} \times \Delta\Phi^{(1)}(\beta_1 + \eta, \beta + \eta)$,

(15b.2)

$$I\,[\rho_1^{\eta_1}\rho_2^{\eta_2}\log(\rho_2), \mathbf{n}] = C^{(00)} \times \{\Delta\Phi^{(1)}(\beta_{12} + \eta_2, \beta_1 + \eta_2) + \Delta\Phi^{(1)}(\beta_1 + \eta, \beta + \eta)\}\,,$$

(15b.3) $I\left[\rho_1^{\eta_1}\log(\rho_1)\rho_2^{\eta_2}\log(\rho_2), \mathbf{n}\right] = C^{(00)} \times \{\Delta\Phi^{(1)}(\beta_1 + \eta, \beta + \eta)^2$

$$+ \Delta\Phi^{(1)}(\beta_{12} + \eta_2, \beta_1 + \eta_2)\Delta\Phi^{(1)}(\beta_1 + \eta, \beta + \eta) + \Delta\Phi^{(2)}(\beta_1 + \eta, \beta + \eta)\}\,,$$

(15b.4)

$$I\,[\rho_1^{\eta_1}\log^2(\rho_1)\rho_2^{\eta_2}, \mathbf{n}] = C^{(00)} \times \{\Delta\Phi^{(1)}(\beta_1 + \eta, \beta + \eta)^2 + \Delta\Phi^{(2)}(\beta_1 + \eta, \beta + \eta)\}$$

(15b.5)

$$I[\rho_1^{\eta_1}\rho_2^{\eta_2}\log^2(\rho_2), \mathbf{n}] = C^{(00)} \times \{ (\Delta\Phi^{(1)}(\beta_{12}+\eta_2, \beta_1+\eta_2) + \Delta\Phi^{(1)}(\beta_1+\eta, \beta+\eta))^2$$

$$+ \Delta\Phi^{(2)}(\beta_{12}+\eta_2, \beta_1+\eta_2) + \Delta\Phi^{(2)}(\beta_1+\eta, \beta+\eta) \},$$

where $C^{(00)} = \frac{\gamma_{\mathbf{n}}}{\Gamma(\beta_{12})} \times \frac{\Gamma(\beta_{12}+\eta_2)}{\Gamma(\beta_1+\eta_2)} \times \frac{\Gamma(\beta_1+\eta)}{\Gamma(\beta+\eta)}$.

Proof: The proof is done for (15b.2). The proofs of the other results follow in a similar manner. The result of Thm. 14b is

$$I[\rho_1^{\eta_1}\rho_2^{\eta_2}, \mathbf{n}] = C^{(00)} = \frac{\gamma_{\mathbf{n}}}{\Gamma(\beta_{12})} \times \frac{\Gamma(\beta_{12}+\eta_2)}{\Gamma(\beta_1+\eta_2)} \times \frac{\Gamma(\beta_1+\eta)}{\Gamma(\beta+\eta)}.$$

Differentiate both sides of this with respect to $\eta_2$. The left-hand side of the differentiated expression is $I[\rho_1^{\eta_1}\rho_2^{\eta_2}\log(\rho_2), \mathbf{n}]$. (The justification of the interchange of the integral and derivative is given in App. C.2.) The derivative of $\frac{\Gamma(\beta_{12}+\eta_2)}{\Gamma(\beta_1+\eta_2)}$ is given by $\Delta\Phi^{(1)}(\beta_{12}+\eta_2, \bar{\beta}+\eta_2)$. The derivative of $\frac{\Gamma(\beta_1+\eta)}{\Gamma(\beta+\eta)}$ is given by $\Delta\Phi^{(1)}(\beta_1+\eta, \beta+\eta)$. Substituting these expressions for the appropriate derivatives in the overall derivative of the right-hand side of the equality above for $C^{(00)}$ gives the claimed result. QED.

## 2d. BAYES ESTIMATORS FOR NON-OVERLAP AND PAIR-WISE OVERLAP TERMS

In this subsection we present those Bayes estimators for the functions of Sec. 1 that can be expressed with non-overlapping and pair-wise overlapping terms. These include the first and second posterior moments of the entropy, mutual information, average, and variance, and the first posterior moments of chi-squared and covariance. The second posterior moments of chi-squared and covariance appear in Sec. 2g. All of the results presented in this section assume a uniform prior.

1. Entropy, $S(\mathbf{p}) = -\Sigma_i p_i \log(p_i)$. This result appears in [1] in a different form and is stated here for completeness.

Theorem 16. If $\mathrm{Re}(\nu_i) > 0$, $i = 1, \ldots, m$ then

(16.1) $E[S(\mathbf{p})|\mathbf{n}] = -\Sigma_i \frac{\nu_i}{\nu} \Delta\Phi^{(1)}(\nu_i + 1, \nu + 1)$,

(16.2) $E[S^2(\mathbf{p})|\mathbf{n}] =$

$$\Sigma_{i \neq j} \frac{\nu_i \nu_j}{\nu(\nu+1)} \times \{\Delta\Phi^{(1)}(\nu_i + 1, \nu + 2)\Delta\Phi^{(1)}(\nu_j, \nu + 2) - \Phi^{(2)}(\nu + 2)\}$$

$$+ \Sigma_i \frac{\nu_i(\nu_i + 1)}{\nu(\nu+1)} \times \{\Delta\Phi^{(1)}(\nu_i + 2, \nu + 2)^2 + \Delta\Phi^{(2)}(\nu_i + 2, \nu + 2)\} .$$

2. Mutual information, $M(\mathbf{p}) = \Sigma_{ij} p_{ij} \log\left(\frac{p_{ij}}{p_{i\cdot}\, p_{\cdot j}}\right)$. In this case the observed counts form a matrix. Define $\bar{\nu}_{ij} \equiv \nu_{i\cdot} + \nu_{\cdot j} - \nu_{ij}$.

Theorem 17. If the $\nu_{ij}$ are non-negative integers $\forall\, ij$ (the integer condition is used only in the simplification of the $\overline{IN}$ term in (17.2); for the other terms it may be relaxed) then

(17.1) $E[M(\mathbf{p})|\mathbf{n}] = \overline{IJ} - \bar{I} - \bar{J}$ where

$$\overline{IJ} = E[\Sigma_{ij} p_{ij} \log(p_{ij})\,|\mathbf{n}] = \Sigma_{ij} \frac{\nu_{ij}}{\nu} \Delta\Phi^{(1)}(\nu_{ij} + 1, \nu + 1),$$

$$\bar{I} = E[\Sigma_i p_{i\cdot} \log(p_{i\cdot})\,|\mathbf{n}] = \Sigma_i \frac{\nu_{i\cdot}}{\nu} \Delta\Phi^{(1)}(\nu_{i\cdot} + 1, \nu + 1),$$

$$\bar{J} = E[p_{\cdot j} \log(p_{\cdot j})\,|\mathbf{n}] = \Sigma_j \frac{\nu_{\cdot j}}{\nu} \Delta\Phi^{(1)}(\nu_{\cdot j} + 1, \nu + 1) \text{, and}$$

(17.2) $E[M^2(\mathbf{p})|\mathbf{n}] = \overline{IJMN} + \overline{IM} + \overline{JN} - 2(\overline{IJM} + \overline{IJN} - \overline{IN})$ where

$$\overline{IJMN} = E[\Sigma_{ijmn} p_{ij} \log(p_{ij})\, p_{mn} \log(p_{mn}) \,|\, \mathbf{n}] =$$

$$\Sigma_{ij}\Sigma_{mn \neq ij} \frac{\nu_{ij}\nu_{mn}}{\nu(\nu+1)} \times \{\Delta\Phi^{(1)}(\nu_{ij}+1, \nu+2)\Delta\Phi^{(1)}(\nu_{mn}+1, \nu+2) - \Phi^{(2)}(\nu+2)\}$$

$$+ \Sigma_{ij} \frac{\nu_{ij}(\nu_{ij}+1)}{\nu(\nu+1)} \times \{\Delta\Phi^{(1)}(\nu_{ij}+2, \nu+2)^2 + \Delta\Phi^{(2)}(\nu_{ij}+2, \nu+2)\}$$

$$\overline{IM} = E[\Sigma_{im} p_{i\cdot} \log(p_{i\cdot})\, p_{m\cdot} \log(p_{m\cdot}) \,|\, \mathbf{n}] =$$

$$\Sigma_{i}\Sigma_{m \neq i} \frac{\nu_{i\cdot}\nu_{m\cdot}}{\nu(\nu+1)} \times \{\Delta\Phi^{(1)}(\nu_{i\cdot}+1, \nu+2)\Delta\Phi^{(1)}(\nu_{m\cdot}+1, \nu+2) - \Phi^{(2)}(\nu+2)\}$$

$$+ \Sigma_{i} \frac{\nu_{i\cdot}(\nu_{i\cdot}+1)}{\nu(\nu+1)} \times \{\Delta\Phi^{(1)}(\nu_{i\cdot}+2, \nu+2)^2 + \Delta\Phi^{(2)}(\nu_{i\cdot}+2, \nu+2)\}$$

(To find $\overline{JN}$ substitute $\nu_{\cdot i}$ for $\nu_{i\cdot}$ and $\nu_{\cdot m}$ for $\nu_{m\cdot}$ in the expression for $\overline{IM}$.)

$$\overline{IJM} = E[\Sigma_{ij}\Sigma_{m} p_{ij} \log(p_{ij})\, p_{m\cdot} \log(p_{m\cdot}) \,|\, \mathbf{n}] =$$

$$\Sigma_{ij}\Sigma_{m \neq i} \frac{\nu_{ij}\nu_{m\cdot}}{\nu(\nu+1)} \times \{\Delta\Phi^{(1)}(\nu_{ij}+1, \nu+2)\Delta\Phi^{(1)}(\nu_{m\cdot}+1, \nu+2) - \Phi^{(2)}(\nu+2)\}$$

$$+ \Sigma_{ij} \frac{\nu_{ij}(\nu_{i\cdot}+1)}{\nu(\nu+1)} \times \{\Delta\Phi^{(1)}(\nu_{i\cdot}+2, \nu+2)^2 + \Delta\Phi^{(1)}(\nu_{ij}+1, \nu_{i\cdot}+1)\Delta\Phi^{(1)}(\nu_{i\cdot}+2, \nu+2)$$

$$+ \Delta\Phi^{(2)}(\nu_{i\cdot}+2, \nu+2)\}$$

(To find $\overline{IJN}$ substitute $\nu_{\cdot m}$ for $\nu_{m\cdot}$ in the expression for $\overline{IJM}$.)

$$\overline{IN} = E[\Sigma_{i}\Sigma_{n} p_{i\cdot} \log(p_{i\cdot})\, p_{\cdot n} \log(p_{\cdot n}) \,|\, \mathbf{n}] =$$

$$\Sigma_{in} \frac{\bar{\nu}_{in}(\bar{\nu}_{in}+1)}{\nu(\nu+1)} \times \{ (\Delta\Phi^{(1)}(\bar{\nu}_{in}+2, \nu+2)^2 + \Delta\Phi^{(2)}(\bar{\nu}_{in}+2, \nu+2)) \times \{1 - \frac{\nu_{i\cdot} + \nu_{\cdot n} - 2\nu_{in}}{\nu} + \frac{(\nu_{i\cdot} - \nu_{in})(\nu_{\cdot n} - \nu_{in})}{\nu(\nu+1)}\}$$

$$+ \Delta\Phi^{(1)}(\bar{\nu}_{in}+2, \nu+2) \times \Sigma_{r=0}^{\infty} \frac{Q_1(r,1)}{r!} \left[ \frac{(\nu_{\cdot n} - \nu_{in})_r}{(\bar{\nu}_{in})_r} \left(1 + \frac{\nu_{i\cdot} - \nu_{in}}{\bar{\nu}_{in} + r}\right) + \frac{(\nu_{i\cdot} - \nu_{in})_r}{(\bar{\nu}_{in})_r} \left(1 + \frac{\nu_{\cdot n} - \nu_{in}}{\bar{\nu}_{in} + r}\right) \right]$$

$$+ \Sigma_{r=0}^{\infty} \Sigma_{s=0}^{\infty} \frac{(\nu_{i\cdot} - \nu_{in})_r (\nu_{\cdot n} - \nu_{in})_s}{(\bar{\nu}_{in})_{r+s}} \frac{Q_1(r,1)}{r!} \frac{Q_1(s,1)}{s!} \}$$

where $Q_1$ is defined in Thm. 15a.

<u>Proof:</u> Write $M(\mathbf{p})$ as $M(\mathbf{p}) = \Sigma_{ij} p_{ij} \log(p_{ij}) - \Sigma_i p_{i\cdot} \log(p_{i\cdot}) - \Sigma_j p_{\cdot j} \log(p_{\cdot j})$.

To prove (17.1) write $E[M(\mathbf{p})|\mathbf{n}] = \overline{IJ} - \bar{I} - \bar{J}$, where $\overline{IJ} \equiv E[\Sigma_{ij} p_{ij} \log(p_{ij}) \mid \mathbf{n}]$ etc. Recall that for the uniform prior $E[F(\mathbf{p})|\mathbf{n}] = I[F(\mathbf{p}), \mathbf{n}] / I[1, \mathbf{n}]$, and apply Thm. 13.1 with $k = 1$, $\eta_u = 1$, $\rho_u = p_{ij}$, $\rho_u = p_{i\cdot}$, $\rho_u = p_{\times j}$ to simplify the numerators in the expressions $\overline{IJ}$, $\bar{I}$, $\bar{J}$ respectively. From Thm. 3, $I[1, \mathbf{n}] = \frac{\gamma_{\mathbf{n}}}{\Gamma(\nu)}$. The results follow by substitution.

To prove (17.2) first square $M(\mathbf{p})$. In a manner similar to the proof of (17.1), find $E[M^2(\mathbf{p})|\mathbf{n}] = \overline{IJMN} + \overline{IM} + \overline{JN} - 2(\overline{IJM} + \overline{IJN} - \overline{IN})$, where each term may be expressed as a ratio of $I[\cdot, \cdot]$ integrals, with $I[1, \mathbf{n}]$ in the denominator. The rest of this proof outlines the calculation of the numerator terms in these ratios.

Apply Thm. 13.2 with $k = 2$, $\eta_u = 1$, $\eta_v = 1$, $\rho_u = p_{ij}$, $\rho_v = p_{mn}$ to find the $ij \neq mn$ terms of $I[\overline{IJMN}|\mathbf{n}]$, and apply Thm. 13.3 with $k = 1$, $\eta_u = 2$, $\rho_u = p_{ij}^2$ to find the $ij = mn$ terms of $I[\overline{IJMN}|\mathbf{n}]$.

Apply Thm. 13.2 with $k = 2$, $\eta_u = 1, \eta_v = 1$, $\rho_u = p_{i\cdot}, \rho_v = p_{m\cdot}$ to find the $i \neq m$ terms of $I[\overline{IM}|\mathbf{n}]$, and apply Thm. 13.3 with $k = 1$, $\eta_u = 2$, $\rho_u = p_{i\cdot}^2$ to find the $i = m$ terms of $I[\overline{IM}|\mathbf{n}]$. To find $I[\overline{JN}|\mathbf{n}]$ substitute $\nu_{\cdot i}$ for $\nu_{i\cdot}$ and $\nu_{\cdot m}$ for $\nu_{m\cdot}$ in the expression for $I[\overline{IM}|\mathbf{n}]$.

Apply Thm. 13.2 with $k = 2$, $\eta_u = 1, \eta_v = 1$, $\rho_u = p_{ij}, \rho_v = p_{m\cdot}$ to find the $i \neq m$ terms of $I[\overline{IJM}|\mathbf{n}]$, and apply Thm. 15b.3 with $\eta_1 = 1, \eta_2 = 1$ and $\rho_1 = p_{i\cdot}, \rho_2 = p_{ij}$ to find the $i = m$ terms of $I[\overline{IJM}|\mathbf{n}]$. To find $I[\overline{IJN}|\mathbf{n}]$ substitute $\nu_{\cdot m}$ for $\nu_{m\cdot}$ in the expression for $I[\overline{IJM}|\mathbf{n}]$.

Apply Thm. 15a.3 with $\eta_1 = 1, \eta_2 = 1$ and $\rho_1 = p_{i\cdot}, \rho_2 = p_{\cdot n}$ to find $I[\overline{IN}|\mathbf{n}]$. QED.

3) Average $A(\mathbf{p}) = \Sigma_{i=1}^{m} p_i X_i$. (Note that for $X_i = \delta_{ij}$, (1) of Thm. 18 below gives the Laplace Law of Succession estimator for $p_j$.)

Theorem 18. If $\mathrm{Re}(\nu_i) > 0 \ \forall i$ then

(18.1) $$E[A(\mathbf{p}) \mid \mathbf{n}] = \Sigma_i \frac{\nu_i}{\nu} X_i .$$

(18.2) $$E[A^2(\mathbf{p}) \mid \mathbf{n}] = \Sigma_{i \neq j} \frac{\nu_i \nu_j}{\nu(\nu+1)} X_i X_j + \Sigma_i \frac{\nu_i(\nu_i+1)}{\nu(\nu+1)} X_i^2 .$$

4) Variance $V(\mathbf{p}) = \Sigma_{i=1}^{m} p_i (X_i - \mu_x)^2$.

Note that $E[V(\mathbf{p}) \mid \mathbf{n}] \neq E[(A(\mathbf{p}) - E[A(\mathbf{p}) \mid \mathbf{n}])^2 \mid \mathbf{n}]$; $\mu_x$ is the true mean, not the expected mean, and $V(\mathbf{p})$ refers to the true variance, not the variance in the estimator $E[A(\mathbf{p}) \mid \mathbf{n}]$.

Theorem 19. If $\mathrm{Re}(\nu_i) > 0 \ \forall i$ then

(19.1) $$E[V(\mathbf{p}) \mid \mathbf{n}] = \Sigma_i \frac{\nu_i(\nu - \nu_i)}{\nu(\nu+1)} X_i^2 - \Sigma_{i \neq j} \frac{\nu_i \nu_j}{\nu(\nu+1)} X_i X_j .$$

(19.2)

$$E[V^2(\mathbf{p}) \mid \mathbf{n}] = \Sigma_{ij} E[p_i p_j \mid \mathbf{n}] X_i^2 X_j^2 - 2\Sigma_{ijk} E[p_i p_j p_k \mid \mathbf{n}] X_i^2 X_j X_k + \Sigma_{ijkl} E[p_i p_j p_k p_l \mid \mathbf{n}] X_i X_j X_k X_l,$$

where the expectations are found by applying Thm. 12.

5) Covariance $C(\mathbf{p}) = \Sigma_{ij} p_{ij} (X_i - \mu_x) (Y_i - \mu_y)$.

<u>Theorem 20.</u> If $\mathrm{Re}(\nu_{ij}) > 0 \ \forall ij$ then

$$E[C(\mathbf{p}) \mid \mathbf{n}] = \Sigma_{ij} X_i Y_j \frac{1}{\nu(\nu+1)} [\nu \nu_{ij} - \nu_{i\cdot} \nu_{\cdot j}],$$

found by applying Thm. 14a.

The second posterior moment of $C(\mathbf{p})$ is given in Thm. 26.

6) Chi-Squared $\chi^2(\mathbf{p}) = \Sigma_{ij} \frac{(p_{ij} - p_{i\cdot} p_{\cdot j})^2}{p_{i\cdot} p_{\cdot j}}$.

<u>Theorem 21.</u> If $\mathrm{Re}(\nu_{ij}) > 0 \ \forall ij$, $\mathrm{Re}(\nu_{i\cdot}) > -1$, and $\mathrm{Re}(\nu_{\cdot j}) > -1$ then

$$E[\chi^2(\mathbf{p}) \mid \mathbf{n}] = -1 + \Sigma_{ij} \frac{(\nu-1)(\nu-2)}{(\nu_{i\cdot} + \nu_{\cdot j} - \nu_{ij} + 1)(\nu_{i\cdot} + \nu_{\cdot j} - \nu_{ij})} \times \Sigma_{m=0}^{\infty} \Sigma_{n=0}^{\infty} \frac{(\nu_{i\cdot} - \nu_{ij})_m (\nu_{\cdot j} - \nu_{ij})_n}{(\nu_{i\cdot} + \nu_{\cdot j} - \nu_{ij} + 2)_{m+n}},$$

found by applying Thm. 14a.

The second posterior moment of $\chi^2(\mathbf{p})$ is given in Thm. 27.

## 2e. NOTATION FOR MULTIPLE OVERLAP INTEGRALS.

The calculation of multiple overlap integrals is not straightforward. The process can be summarized as follows. First T transforms are applied and then Thms. 9.1 and 9.2 are used, leaving a convolution product of m terms. Each term in this product has the form $\tau^{\alpha_k - 1}\Pi_{i=1}^{n} e^{-\tau a_i t_i}$, $k = 1, \ldots, m$, with the $a_i$'s taking values in $\{0, 1\}$, the $t_i$'s the T transform variables, $\tau$ the convolution variable, and the $\alpha_k$'s constants (see Thm. 14a for an example). The upper index on the product, $n$, is determined by the overlap structure.

Since the convolution operation is both commutative and associative, the convolutions may be done in any order. However, if the convolutions are done in random order, it is quite easy to arrive at an expression for which the inverse T transforms cannot be evaluated in closed form. Further, given any particular ordering of the convolution operations, much bookkeeping needs to be done to actually find the result. Thus, it is important to have a method for quickly determining the salient aspects of any convolution ordering (whether or not the particular order chosen for the convolutions leads to an invertible expression), without actually having to do the convolutions.

In order to facilitate this, *Convolution Form* (CF) notation is now introduced. This notation captures the relevant aspects of the expressions involved and provides a guide for the rapid calculation of multiple overlap convolution integrals. First, we show how to translate convolution product terms into this notation. Then we use the CF notation to state the relevant algebraic properties of convolutions. It is not expected that the justifications for these properties should be transparent to the reader. In fact, it is not even expected that the reader will have a complete and formal understanding of CF notation; in the interests of simplicity, CF's are presented here simply as a useful algebraic framework for performing calculations. The reader interested in the formal details is directed to App. G.

Write each expression $\tau^{\alpha_k - 1}\Pi_{i=1}^{n} e^{-\tau a_i t_i}$ appearing in the convolution product (the T transformed integral mentioned two paragraphs ago) in CF notation as the CF $(\mathbf{a};\mathbf{0})$, where **a** is the $n$- vector of constants $a_i$, and where $\mathbf{0}$ is the zero $n$-vector. In writing $(\mathbf{a}, 0)$, the $t$, $\tau$ and $\alpha_k$ dependence is assumed.

The second component $\mathbf{b}$ of the general CF $(\mathbf{a};\mathbf{b})$ may be nonzero (see App. G). However, for the moment there is no need to present the full definition of the CF's encompassing nonzero $\mathbf{b}$; for now $\mathbf{b}$ should be taken simply as an algebraic object with certain useful properties (these properties are formally justified in App. G). The reader should be aware though that in general, for nonzero $\mathbf{b}$, $(\mathbf{a};\mathbf{b})$ represents a set of expressions; i.e. $(\mathbf{a};\mathbf{b})$ does not uniquely specify a single expression. (I.e., it does not uniquely specify a function of t, $\tau$ and the $\alpha_k$; see App. G.)

We represent the convolution of two CF's using brackets, so that the convolution of the CF's $(\mathbf{a};\mathbf{b})$ and $(\mathbf{c};\mathbf{d})$ is given by $[(\mathbf{a};\mathbf{b}), (\mathbf{c};\mathbf{d})]$. The general algebraic rule for translating this convolution into a CF is

$$[(\mathbf{a};\mathbf{b}), (\mathbf{c};\mathbf{d})] \subset (\mathbf{c}\ ;\ \mathbf{b} \vee \mathbf{d} \vee \mathrm{nz}(\mathbf{c} - \mathbf{a})), \tag{13}$$

where: i) "$\vee$" is the vector-or operator with $[\mathbf{a} \vee \mathbf{b}]_i = 1$ if $a_i = 1$ or $b_i = 1$, $0$ otherwise; ii) "-" is the vector-difference operator with $[\mathbf{a} - \mathbf{c}]_i = a_i - c_i$, and iii) "nz" is the vector-valued nonzero operator, with $[\mathrm{nz}(\mathbf{a})]_i = 1$ if $a_i \neq 0$, $0$ otherwise. We use the symbol "$\subset$" because Eq. (13) is a relationship between two *sets* of expressions; it means that the convolution product *may be written in the form* $(\mathbf{c}\ ;\ \mathbf{b} \vee \mathbf{d} \vee \mathrm{nz}(\mathbf{c} - \mathbf{a}))$ (see App. G).

Now we state the algebra of CF's under inverse T transforms. Let $(\mathbf{a}\ ;\ \mathbf{b})$ be any CF with $a_i = 0 = b_i$, for some $i \in \{1, \ldots, m\}$. Further, let $c \neq 0$. Then we have

$$T_i^{-1}[(c\mathbf{e}_i + \mathbf{a}\ ;\ \mathbf{b})(x)] \subset (\mathbf{a}\ ;\ \mathbf{b}), \text{ and } T_i^{-1}[(c\,\mathbf{e}_i + \mathbf{a}\ ;\ \mathbf{e}_i + \mathbf{b})(x)] \subset (\mathbf{a}\ ;\ \mathbf{b}), \tag{14}$$

where the inverse transform is with respect to x, and $[\mathbf{e}_i]_j \equiv \delta_{ij}$, the Kronecker-delta function.

In the theorems that follow, any vector occurring as an argument of a CF has all of its components equal to either $0$ or $1$. Such vectors will be represented by writing a shorthand list of the indices of the non-zero values only. For example, the vector $(1, 0, 1, 1)$ will be represented by the list $\{1, 3, 4\}$. Finally, the $\{\ \}$'s will be dropped from the lists since they are burdensome, and empty lists will be represented with the dash symbol “-”. The result of a successful ordering of convolutions and inversions is the CF $(-;-)$, i.e. our goal is to find an ordering of the convolution operations such that the repeated applications of Eqs. (13) and (14) result in $(-;-)$. (See App. G for the justification of this being the condition that allows for the closed form evaluation of the inverse T transforms.)

## 2f. MULTIPLE OVERLAP INTEGRALS.

The CF identities (Sec. 2e, Eqs. (13) and (14)) allow the efficient calculation of the needed multiple overlap integrals, as we demonstrate in Thms. 22 and 23. Theorems 24 and 25 apply Thms. 22 and 23 to evaluate the integrals appearing in the Bayes estimators for the second moments of covariance and chi-squared. In section 2g we use these results to write closed form expressions for these moments.

Theorem 22 concerns convolutions in which there are three subsets of indices $\sigma_1, \sigma_2, \sigma_3$ with i) $\sigma_{12} \neq \emptyset$ and $\sigma_{23} \neq \emptyset$, ii) $\sigma_{13} = \emptyset$, and iii) none of the intersections are equal to any of the subsets $\sigma_1, \sigma_2, \sigma_3$. Theorem 23 concerns convolutions in which there are four subsets of indices, where i) the only non-empty $\sigma_{ij}$ with $i \neq j$ are $\sigma_{12}, \sigma_{23}, \sigma_{34}, \sigma_{41}$, and where ii) none of the intersections are equal to any of the subsets $\sigma_1, \sigma_2, \sigma_3, \sigma_4$.

Overlap structures of these types occur in the integrals for the second moments of covariance and chi-squared. Many mathematically equivalent forms of the results of Thms. 22 and 23 are possible; we present the result of one particular calculation in each case and note that equating different forms leads to hypergeometric identities along the lines of those discussed in footnotes 1 and 2.

In what follows “Terms 1 and 12”, for example, indicates that the terms $e^{-t_1 p} p^{\alpha_1 - 1}$ and $e^{-(t_1 + t_2)} p^{\alpha_{12} - 1}$ are being convolved. Define the shorthand notations $(\alpha, \beta, \gamma) \equiv \frac{\Gamma(\alpha)\Gamma(\beta)\Gamma(\gamma)}{\Gamma(\alpha + \beta + \gamma)}$, $(\alpha, \beta) \equiv \frac{\Gamma(\alpha)\Gamma(\beta)}{\Gamma(\alpha + \beta)}$. As in App. A, the symbol $(\alpha)_i \equiv \frac{\Gamma(\alpha + i)}{\Gamma(\alpha)}$.

Theorem 22 states the result of a derivation that treats subsets 1 and 3 symmetrically from the outset. For future use, in Thm. 22 the expression being evaluated is defined simply as $I_3$, suppressing the various $\alpha$ and $\eta$ arguments. (The inverse T transforms turn $t$'s into $\eta$'s.)

Theorem 22. If $\mathrm{Re}(\alpha_1) > 0$, $\mathrm{Re}(\alpha_{12}) > 0$, $\mathrm{Re}(\alpha_2) > 0$, $\mathrm{Re}(\alpha_{23}) > 0$, $\mathrm{Re}(\alpha_3) > 0$, and $\alpha \equiv \alpha_1 + \alpha_{12} + \alpha_2 + \alpha_{23} + \alpha_3$, then

$$I_3 \equiv T_1^{-1} T_2^{-1} T_3^{-1} [e^{-t_1 p} p^{\alpha_1 - 1} \otimes e^{-(t_1 + t_2) p} p^{\alpha_{12} - 1} \otimes e^{-t_2 p} p^{\alpha_2 - 1} \otimes e^{-(t_2 + t_3) p} p^{(\alpha_{23} - 1)} \otimes e^{-t_3 p} p^{\alpha_3 - 1}]$$

$$= \tau^{\alpha + \eta - 1} (\alpha_1, \alpha_{12}) (\alpha_3, \alpha_{23}) (\alpha_1 + \alpha_{12} + \eta_1, \alpha_3 + \alpha_{23} + \eta_3, \alpha_2) \times {}_{2,2,1}F_{1,1,1} [(\alpha_1, \alpha_1 + \alpha_{12} + \eta_1), (\alpha_3, \alpha_3 + \alpha_{23} + \eta_3), -\eta_2; \alpha_1 + \alpha_{12}, \alpha_3 + \alpha_{23}, \alpha + \eta_1 + \eta_3; 1, 1]$$

Proof: The following derivation applies Eqs. (13) and (14) repetitively to demonstrate that a particular ordering of the convolutions results in the sucessfully inverted CF $(-;-)$.

| | | | |
|---|---|---|---|
| Terms 1 and 12: | $[(1;-), (1,2;-)]$ | $\subset (1,2;2)$ | |
| | $T_1^{-1}[(1,2;2)]$ | $\subset (2;2)$ | (a) |
| Terms 3 and 23: | $[(3;-), (2,3;-)]$ | $\subset (2,3;2)$ | |
| | $T_3^{-1}[(2,3;2)]$ | $\subset (2;2)$ | (b) |
| Results (a) and (b): | $[(2;2), (2;2)]$ | $\subset (2;2)$ | (c) |
| Res. (c) and term 2: | $[(2;2), (2;-)]$ | $\subset (2;2)$ | (d) |
| Invert (d): | $T_2^{-1}[(2;2)]$ | $\subset (-;-)$. | |

Now, carry out the convolutions and inverses in the above order with the appropriate terms substituted as indicated in the left column to find

$$I_3 = \tau^{\alpha+\eta-1}\Sigma_{i=0}^{\infty}\Sigma_{j=0}^{\infty}\frac{1}{i!j!}(-\eta_2)_{i+j}(\alpha_1+i,\alpha_{12})(\alpha_3+j,\alpha_{23})$$

$$\times(\alpha_1+\alpha_{12}+i+\eta_1,\alpha_3+\alpha_{23}+j+\eta_3,\alpha_2)\ .$$

Rewriting this using hypergeometric notation (see App. A) gives the desired result. QED.

Theorem 23 presents the results for the relevant quadruple overlap sum convolution. As in Thm. 22, here we use the notation $(\alpha,\beta,\gamma)\equiv\frac{\Gamma(\alpha)\Gamma(\beta)\Gamma(\gamma)}{\Gamma(\alpha+\beta+\gamma)}$ and $(\alpha,\beta)\equiv\frac{\Gamma(\alpha)\Gamma(\beta)}{\Gamma(\alpha+\beta)}$. In order to simplify the resulting expressions, we do not express the result in hypergeometric notation. For future use, in Thm. 23 the expression being evaluated is defined simply as $I_4$, suppressing the various $\alpha$ and $\eta$ arguments.

Theorem 23. If $\mathrm{Re}(\alpha_1)>0$, $\mathrm{Re}(\alpha_{12})>0$, $\mathrm{Re}(\alpha_2)>0$, $\mathrm{Re}(\alpha_{23})>0$, $\mathrm{Re}(\alpha_3)>0$, $\mathrm{Re}(\alpha_{34})>0$, $\mathrm{Re}(\alpha_4)>0$, and $\alpha\equiv\alpha_1+\alpha_{12}+\alpha_2+\alpha_{23}+\alpha_3+\alpha_{34}+\alpha_4+\alpha_{41}$ then

$$I_4\equiv T_1^{-1}T_2^{-1}T_3^{-1}T_4^{-1}[e^{-t_1p}p^{\alpha_1-1}\otimes e^{-(t_1+t_2)p}p^{\alpha_{12}-1}\otimes e^{-t_2p}p^{\alpha_2-1}$$

$$\otimes e^{-(t_2+t_3)p}p^{\alpha_{23}-1}\otimes e^{-t_3p}p^{\alpha_3-1}\otimes e^{-(t_3+t_4)p}p^{\alpha_{34}-1}\otimes e^{-t_4p}p^{\alpha_4-1}]$$

$$=\tau^{\alpha+\eta-1}\Sigma_{i,j,m,n,p,q=0}^{\infty}(-1)^{i+j+m+n}\frac{1}{i!j!m!n!p!q!}(-\eta_2)_{j+m+p}(-\eta_4)_{i+n+q}$$

$$\times(\alpha_{41}+i,\alpha_1,\alpha_{12}+j)(\alpha_{23}+m,\alpha_3,\alpha_{34}+n)$$

$$\times(\alpha_{41}+\alpha_1+\alpha_{12}+\eta_1+i+j,\alpha_{23}+\alpha_3+\alpha_{34}+\eta_3+m+n)$$

$$\times(\alpha-\alpha_2-\alpha_4+\eta_1+\eta_3+i+j+m+n+p,\alpha_2)$$

$$\times(\alpha-\alpha_4+\eta_1+\eta_2+\eta_3+i+n+q,\alpha_4)\ .$$

Proof: The following derivation applies Eqs. (13) and (14) repetitively to demonstrate that a particular ordering of the convolutions results in the successfully inverted CF (-;-). Initially, con-volutions are done on two non-overlapping groups - the group depending on $t_1$ and the group depending on $t_3$. Then the inverses $T_1^{-1}$ and $T_3^{-1}$ are taken and the remaining convolutions are done with care taken to create expressions that have simple T transform inverses.

For the $t_1$ group (terms 41, 1, 12),

| | | | |
|---|---|---|---|
| Terms 41 and 1: | $[\,(4, 1;-)\,,\ (1;-)\,]$ | $\subset (1;4)$ | (a) |
| Term 12 and Res. (a): | $[\,(1, 2;-)\,,\ (1;4)\,]$ | $\subset (1;2, 4)$ | |
| | $T_1^{-1}\,[\,(1;2, 4)\,]$ | $\subset (-;2, 4)$ | (b) |

Similarly for the $t_3$ group (terms 23, 3, 34):$\subset (-;2, 4)$ (c)

| | | | |
|---|---|---|---|
| Res. (b) and (c): | $[\,(-;2, 4)\,,\ (-;2, 4)\,]$ | $\subset (-;2, 4)$ | (d) |
| Res. (d) and term 2: | $[\,(-;2, 4)\,,\ (2;-)\,]$ | $\subset (2;2, 4)$ | |
| | $T_2^{-1}\,[\,(2;2, 4)\,]$ | $\subset (-;4)$ | (e) |
| Res. (e) and term 4: | $[\,(-;4)\,,\ (4;-)\,]$ | $\subset (4;4)$ | |
| | $T_4^{-1}\,[\,(4;4)\,]$ | $\subset (-;-)$ . | |

Now, carry out the convolutions and inverses in the above order with the appropriate terms substituted as indicated in the left column to find the desired result. QED.

Theorems 24 and 25 utilize Thms. 22 and 23 respectively to find the needed multiple overlap integrals. They are given without proof, as they follow immediately from these theorems, Thm. 9.1, and induction.

Theorem 24. Assume the overlap structure relevant to Thm. 22, with $C_3$ defined as the $\tau$-independent factor of the $I_3$ defined in Thm. 22. (I.e., $I_3 = \tau^{\alpha+\eta-1}C_3$.) Define $\bar{\beta} \equiv \beta - \beta_{1+2+3}$. If $\mathrm{Re}\,(\beta_u + \eta_u) > 0$ for $u = 1, 2, 3$, then

$$I\,[\rho_1^{\eta_1}\rho_2^{\eta_2}\rho_3^{\eta_3}, \mathbf{n}] = \frac{\Gamma(\beta_{1+2+3}+\eta)\Gamma(\bar{\beta})}{\Gamma(\beta+\eta)} \times C_3\,[\beta_{1-12}, \beta_{12}, \beta_{2-12-23}, \beta_{23}, \beta_{3-23}; \eta_1, \eta_2, \eta_3]\,.$$

Theorem 25. Assume the overlap structure relevant to Thm. 23, with $C_4$ defined as the $\tau$-independent factor of the $I_4$ defined in Thm. 23. (I.e., $I_4 = \tau^{\alpha+\eta-1}C_4$.) Define $\bar{\beta} \equiv \beta - \beta_{1+2+3+4}$. If $\mathrm{Re}\,(\beta_u + \eta_u) > 0$ for $u = 1, 2, 3, 4$, then

$$I\,[\rho_1^{\eta_1}\rho_2^{\eta_2}\rho_3^{\eta_3}\rho_4^{\eta_4}, \mathbf{n}] = \frac{\Gamma(\beta_{1+2+3+4}+\eta)\Gamma(\bar{\beta})}{\Gamma(\beta+\eta)}$$

$$\times C_4\,[\beta_{1-12}, \beta_{12}, \beta_{2-12-23}, \beta_{23}, \beta_{3-23-34}, \beta_{34}, \beta_{4-34-41}, \beta_{41}; \eta_1, \eta_2, \eta_3, \eta_4]\,.$$

## 2g. BAYES ESTIMATORS - MULTIPLE OVERLAP TERMS.

In this subsection we complete the presentation of the results for the Bayes estimators (see Sec. 2d) by giving the Bayes estimators with uniform prior for the second powers of covariance and chi-squared. Since the complete description of these results is *quite* lengthy, we present them in recipe form; at this point the reader should be able to make the needed substitutions.

5) Covariance $C(\mathbf{p}) = \Sigma_{ij} p_{ij} (X_i - \mu_x)\,(Y_i - \mu_y)\,.$

Theorem 26. If $\mathrm{Re}\,(\nu_{ij}) > 0\ \forall\, ij$ then

$$E\,[C^2(\mathbf{p}) \mid \mathbf{n}] = \Sigma_{ijkl} E\,[p_{ij}p_{kl} \mid \mathbf{n}]\, X_i Y_j X_k Y_l - 2\Sigma_{ijkl} E\,[p_{ij}p_{k\cdot}p_{\cdot l} \mid \mathbf{n}]\, X_i Y_j X_k Y_l$$
$$+ \Sigma_{ijkl} E\,[p_{i\cdot}p_{\cdot j}p_{k\cdot}p_{\cdot l} \mid \mathbf{n}]\, X_i Y_j X_k Y_l,$$

where $E\,[p_{ij}p_{kl} \mid \mathbf{n}]$ is found by applying Thm. 12, $E\,[p_{ij}p_{k\cdot}p_{\cdot l} \mid \mathbf{n}]$ is found by applying Thm. 14a, and $E\,[p_{i\cdot}p_{\cdot j}p_{k\cdot}p_{\cdot l} \mid \mathbf{n}]$ is either a single, double, or quadruple overlap term found by applying Thms. 14a, 24, and 25: if $i = k, j = l$, then apply Thm. 14a; if $i = k, j \neq l$ or $i \neq k, j = l$, then apply Thm. 24; if $i \neq k, j \neq l$ then apply Thm. 25.

6) Chi-Squared $\chi^2(\mathbf{p}) = \Sigma_{ij} \frac{(p_{ij} - p_{i.}p_{.j})^2}{p_{i.}p_{.j}}$.

Theorem 27. If $\mathrm{Re}(\nu_{ij}) > 0 \; \forall ij$, $\mathrm{Re}(\nu_{i.}) > -1$ and $\mathrm{Re}(\nu_{.j}) > -1$ then

$$E[(\chi^2(\mathbf{p}))^2 \mid \mathbf{n}] = \Sigma_{ijkl} E\left[\frac{p_{ij}^2 p_{kl}^2}{p_{i.}p_{.j}p_{k.}p_{.l}} \mid \mathbf{n}\right] - 2E[\chi^2(\mathbf{p}) \mid \mathbf{n}] - 1,$$

where $E[\chi^2(\mathbf{p}) \mid \mathbf{n}]$ is given in Thm. 21 and where $E\left[\frac{p_{ij}^2 p_{kl}^2}{p_{i.}p_{.j}p_{k.}p_{.l}} \mid \mathbf{n}\right]$ is either a single, double, or quadruple overlap term found by applying Thms. 14a, 24, and 25: if $i = k, j = l$, then apply Thm. 14a; if $i = k, j \neq l$ or $i \neq k, j = l$, then apply Thm. 24; if $i \neq k, j \neq l$ then apply Thm. 25.

## 3. EXTENSION OF THE CLASS OF CALCULABLE PRIORS

The calculations of this paper and [1] are done under the assumption of a uniform prior $P(\mathbf{p})$. However they also apply essentially unchanged when certain other priors are used. Here we briefly discuss applying the calculations of these papers to cases where the prior is not uniform. As specific examples of how the calculations of these papers are modified when the prior is not uniform, we consider priors of the form $P(\mathbf{p}) \propto \Delta(\mathbf{p})\Pi_{i=1}^m p_i^{r_i}$ (the Dirichlet priors are the subset of these with all $r_i$ equal) and the entropic prior $P(\mathbf{p}) \propto \Delta(\mathbf{p})\, e^{\alpha S(\mathbf{p})}$.

Define $\bar{P}(\mathbf{p})$ implicitly through $P(\mathbf{p}) = \Delta(p)\bar{P}(\mathbf{p})$. ($\bar{P}(\mathbf{p})$ is the non-delta function part of $P(\mathbf{p})$). The uniform prior has a constant $\bar{P}(\mathbf{p})$. Even if $P(\mathbf{p})$ is not uniform, it is often the case that $I[F(\mathbf{p})\bar{P}(\mathbf{p}), \mathbf{n}] = \int d\mathbf{p} F(\mathbf{p}) P(\mathbf{p}) \Pi_{i=1}^m p_i^{n_i}$ and $I[\bar{P}(\mathbf{p}), \mathbf{n}] = \int d\mathbf{p} P(\mathbf{p}) \Pi_{i=1}^m p_i^{n_i}$ are of the form of integrals evaluated in these papers. (In this section all integrals are over $\mathbf{p}$'s with nonnegative com-ponents). In these cases we can evaluate the Bayes estimator for $F(\mathbf{p})$ with prior $P(\mathbf{p})$ since $E[F(\mathbf{p}) \mid \mathbf{n}] = I[F(\mathbf{p})\bar{P}(\mathbf{p}), \mathbf{n}] / I[\bar{P}(\mathbf{p}), \mathbf{n}]$.

For example, the Bayes estimator for $F(\mathbf{p})$ with prior $P(\mathbf{p})$ can often be evaluated using the calculations of these papers when $P(\mathbf{p}) \propto \Delta(\mathbf{p})\Pi_{i=1}^{m} f_i(p_i)$. In particular consider priors of the form $P(\mathbf{p}) \propto \Delta(\mathbf{p})\Pi_{i=1}^{m} p_i^{r_i}$ with $\mathrm{Re}(r_i) > -1$, $i = 1, \ldots, m$. (Dirichlet priors have $r_i = r$ for all $i$.) When $P(\mathbf{p})$ is of this form the Bayes estimator for $F(\mathbf{p})$ with prior $P(\mathbf{p})$ is given by

$$E[F(\mathbf{p}) \mid \mathbf{n}] = \frac{I[F(\mathbf{p}), \mathbf{n} + \mathbf{r}]}{I[1, \mathbf{n} + \mathbf{r}]}. \tag{15}$$

As another example, consider the class of priors with $\bar{P}(\mathbf{p})$ represented by a taylor series converging everywhere in the domain of $\mathbf{p}$. Using this taylor series representation expand both $I[F(\mathbf{p})\bar{P}(\mathbf{p}), \mathbf{n}]$ and $I[\bar{P}(\mathbf{p}), \mathbf{n}]$ into infinite sums of integrals. If all of the integrals are of a form evaluated in this paper the we can find $E[F(\mathbf{p}) \mid \mathbf{n}]$ for the prior $P(\mathbf{p})$.

For example, consider the entropic prior $P(\mathbf{p}) \propto \Delta(\mathbf{p})\ e^{\alpha S(\mathbf{p})}$. In some applications the entropy $S(\mathbf{p})$ is taken to be $S(\mathbf{p}) \equiv -\Sigma_{i=1}^{m} p_i \log(p_i)$, where $\alpha$ is some constant, while in image processing applications $S(\mathbf{p})$ is often defined as $S(\mathbf{p}) \equiv \Sigma_{i=1}^{m}\left((p_i - m_i) - p_i \log\left(\frac{p_i}{m_i}\right)\right)$, where $\mathbf{m}$ is known as the "model" [11]. In either case, $\bar{P}(\mathbf{p})$ may be expanded in the series

$$\bar{P}(\mathbf{p}) = \Sigma_{i=0}^{\infty} \frac{\alpha^i S^i(\mathbf{p})}{i!}, \tag{16}$$

Whenever the products $S^i(\mathbf{p})F(\mathbf{p})$, $i = 0, 1, \ldots$, are of the form of some function integrated in these papers then closed form results (up to series truncations) for the Bayes estimator for $F(\mathbf{p})$ with an entropic prior are available. An application of these ideas appears in [12], where they are used to calculate the normalization constant of the entropic prior

## ACKNOWLEDGMENTS

We would like to thank Alan Lapedes for suggesting the use of the Mellin (inverse-T) transform. We would also like to thank the Center for Nonlinear Studies and Theoretical Division, both of Los Alamos National Laboratory, for their partial support during part of this work. This work was supported in part by the United States Department of Energy under contract number W-7405-ENG-36. David H. Wolpert was supported in part by National Library of Medicine grant number NLM F37-LM00011 and SFI core funding.

## APPENDICES

## A. HYPERGEOMETRIC FUNCTION NOTATION

Here we present the notation for the hypergeometric functions used in this paper. Paralleling Lebedev [13], let **a** and **b** be vectors of dimensions p and q respectively. Define $(\lambda)_k \equiv \Gamma(\lambda + k) / \Gamma(\lambda)$. Define the single summation hypergeometrics ${}_pF_q$ by

$$ {}_pF_q[\mathbf{a};\mathbf{b};\tau] \equiv \Sigma_{i=0}^{\infty}\left(\frac{\Pi_{\alpha=1}^{p}(a_\alpha)_i}{\Pi_{\beta=1}^{q}(b_\beta)_i}\right)\frac{\tau^i}{i!}. \qquad \text{(A.1)} $$

An example of a single summation hypergeometric is ${}_1F_1[\alpha;\beta;\tau]$, which has the integral representation for $\beta > \alpha > 0$ [13]

$$ {}_1F_1[\alpha;\beta;\tau] = \frac{\Gamma(\beta)}{\Gamma(\alpha)\Gamma(\beta-\alpha)}\int_0^1 dx\ e^{\tau x}\ x^{\alpha-1}(1-x)^{\beta-\alpha-1}. \qquad \text{(A.2)} $$

Now, given vectors $\mathbf{a}^1, \mathbf{a}^2, \mathbf{a}^{12}, \mathbf{b}^1, \mathbf{b}^2, \mathbf{b}^{12}$ of dimensions $p_1, p_2, p_{12}, q_1, q_2, q_{12}$ respectively, define the double summation hypergeometrics

$$ {}_{p_1, p_2, p_{12}}F_{q_1, q_2, q_{12}}[\mathbf{a}^1, \mathbf{a}^2, \mathbf{a}^{12};\mathbf{b}^1, \mathbf{b}^2, \mathbf{b}^{12};\tau_1, \tau_2] \equiv $$

$$ \Sigma_{i=0}^{\infty}\Sigma_{j=0}^{\infty}\frac{(\Pi_{\alpha_1=1}^{p_1}(a_{\alpha_1}^1)_i)\ (\Pi_{\alpha_2=1}^{p_2}(a_{\alpha_2}^2)_j)\ (\Pi_{\alpha_{12}=1}^{p_{12}}(a_{\alpha_{12}}^{12})_{i+j})}{(\Pi_{\beta_1=1}^{q_1}(b_{\beta_1}^1)_i)\ (\Pi_{\beta_2=1}^{q_2}(b_{\beta_2}^2)_j)\ (\Pi_{\beta_{12}=1}^{q_{12}}(b_{\beta_{12}}^{12})_{i+j})}\frac{\tau_1^i}{i!}\frac{\tau_2^j}{j!} \qquad \text{(A.3)} $$

In writing arguments of F's, vectors will be denoted by a list of the elements, e.g., $\mathbf{c} = (c_1, \ldots, c_k)$. However, when listing the components of a 1-dimensional vector the parentheses will be dropped. Further, when any of the p or q subscripts are zero (which corresponds to an empty argument for that position), the empty vector argument of the hypergeometric will simply be omitted from the list of arguments.

## B. TRANSFORMS

In App. B.1. we discuss the T transform. In App. B.2. we discuss the Z transform.

### B.1. THE T TRANSFORM

In order to calculate averages of the form $E[(\rho^{\eta})(\log(\rho^{\eta}))\mid \mathbf{n}]$ (with $\rho \equiv \Sigma_i p_{ij}$) and similar averages, the identity

$$\Gamma(-\eta) = \int_0^{\infty} u^{-\eta-1} e^{-u}\, du,\ \mathrm{Re}(-\eta) > 0 \qquad \text{(B.1)}$$

for the gamma function is needed. (Here $-\eta$ has been used instead of $\eta$ in order to simplify the following.) Make the change of variables $u = \rho t$. With $\rho > 0$, independent of $t$, we find

$$\rho^{\eta} = \frac{1}{\Gamma(-\eta)} \int_0^{\infty} t^{-\eta-1} e^{-\rho t}\, dt,\ \mathrm{Re}(-\eta) > 0,\ \rho > 0. \qquad \text{(B.2)}$$

Define the operator $T^{-1}$ by $T^{-1}[F(\cdot)](\eta) \equiv \frac{1}{\Gamma(-\eta)} \int_0^{\infty} t^{-\eta-1} F(t)\, dt,\ \mathrm{Re}(-\eta) > 0$, and define the transform T by $T[T^{-1}[F(\cdot)]] = F(\cdot)$. As defined, the transform T is closely related to the Mellin transform [14] (it is an inverse-Mellin transform) and we rely on this similarity to establish the conditions for the existence of the transform and its inverse. Of interest in this work is the following: For $\rho$ independent of $t$, the functions $\rho^{\eta}$ and $e^{-\rho t}$ form a transform pair. That is,

$$T[\rho^{\eta}](t) = e^{-\rho t} \text{ and } T^{-1}[e^{-\rho t}](\eta) = \rho^{\eta}. \qquad \text{(B.3)}$$

As an example of the use of the transform T, when finding $E[\rho^{\eta}\mid \mathbf{n}]$, where $\rho$ is a sum of a subset of the $p_i$'s and $\mathrm{Re}(\eta) < 0$, the T transform will be taken of $E[\rho^{\eta}\mid \mathbf{n}]$ with respect to $\eta$, which substitutes $e^{-\rho t}$ for $\rho^{\eta}$:

$$T[E[\rho^\eta \mid \mathbf{n}]](t) = T\left[\int d\mathbf{p}\ P(\mathbf{p} \mid \mathbf{n})\ \rho^\eta\right] = \int d\mathbf{p}\ P(\mathbf{p} \mid \mathbf{n})\ T[\rho^\eta] = \int d\mathbf{p}\ P(\mathbf{p} \mid \mathbf{n})\ e^{-\rho t}. \tag{B.4}$$

See App. C.1 for the justification of the commutation of the integrals. After the transform has been done, often the last integral may be computed in closed form (see Sec. 2) and then the $T^{-1}$ transform may be applied.

## B.2. THE Z TRANSFORM

Let $f(\mathbf{n})$ be any function that factors as $f(\mathbf{n}) = \Pi_{i=1}^{m} f_i(n_i)$. For such functions, the Z transform $Z[f](z) \equiv \Sigma_{n=0}^{\infty} f(n) z^n$ is useful in simplifying calculations involving sums $\Sigma_{\mathbf{n}} f(\mathbf{n})$, where the summation extends over all **n** having non-negative integer components and $\Sigma_i n_i = N$. Define the discrete convolution product of two functions $g$ and $h$ by $(g \otimes h)(n) \equiv \Sigma_{i=0}^{n} g(i) h(n-i)$. (Note that $\otimes$ is both commutative and associative, so that the order that the convolutions are taken in is irrelevant, justifying the use of the above notation when several functions are involved.)

The Z transform convolution theorem may be thought of as a discretized form of the Laplace convolution theorem (see Thm. 2).

<u>Theorem B.1:</u> If $F(N) \equiv \Sigma_{\mathbf{n}} f(\mathbf{n})$ and $f(\mathbf{n}) = \Pi_{i=1}^{m} f_i(n_i)$ then $F(N) = (\otimes_{i=1}^{m} f_i)(N)$ and $Z[F](z) = \Pi_{i=1}^{m} Z[f_i](z)$, for all z such that $Z[f_i](z)$, $i = 1, \ldots, m$, converges.

<u>Proof:</u> For $m = 2$ we have $F(N) = (f_1 \otimes f_2)(N)$ and the Z transforms of $f_1$ and $f_2$ are given by $Z[f_i](z) = \Sigma_{n=0}^{\infty} f_i(n) z^n$, $i = 1, 2$, respectively. For $z$ within the radii of convergence of both of these power series, we have (after collecting terms having the same power of $z$)

$$Z[f_1](z) \times Z[f_2](z) = \Sigma_{n=0}^{\infty} z^n \Sigma_{i=0}^{n} f_1(i) f_2(n-i),$$

The right-hand side is immediately seen to be $Z[F](z)$. The result for arbitrary $m$ follows by induction. QED.

Note that due to the uniqueness of power series representations, inverses of Z transforms exist on the nonnegative integers.

## C. COMMUTING LINEAR OPERATORS

In App. C.1. we discuss the interchange of integrals. In App. C.2. we discuss the interchange of derivatives and integrals.

### C.1. COMMUTING TWO INTEGRALS

Interchanging the integrals appearing in these papers as the **p** integral and the T transform integral is possible due to Fubini's theorem [15], which justifies the interchange of uncoupled integrations (region of integration of either integral does not depend on the other integral's parameters) when the double integral exists.

### C.2. COMMUTING INTEGRALS AND DERIVATIVES

Consider differentiating the integral $\int F(x, t)\, dx$ with respect to $t$. Theorem C.1 generalizes Thm. 9.42 of [10] and establishes conditions general enough to allow the commutation of the derivative and integral for the functions $F(x, t)$ appearing in this paper. Define $D_2F(x, t)$ to be the partial derivative of $F$ with respect to its second argument, evaluated at $(x, t)$.

Theorem C.1: If (1) $F(x, t)$ and $D_2F(x, t)$ are defined for $(x, t) \in \Delta_x \times \Delta_t$,

where $\Delta_x = (0, \infty)$, and where $\Delta_t$ is convex,

(2) $\int_0^\infty F(x, t)dx$ exists $\forall t \in \Delta_t$,

(3) $\forall \varepsilon > 0$ and $b > 0$, $\exists\ f(x)$ with $f(x) > 0$ for $x \in \Delta_x$, and $\delta > 0$, $\ni$

$\int_b^\infty f(x)dx < \varepsilon$ and

$\forall x \geq b, \forall s, t \in \Delta_t, |t - s| < \delta \Rightarrow |D_2F(x, t) - D_2F(x, s)| < f(x)$

then $D_2\int_0^\infty F(x, t)dx = \int_0^\infty D_2F(x, t)dx$ on $\Delta_x \times \Delta_t$.

<u>Proof:</u> Let $\phi(s, t) \equiv \frac{F(x, t) - F(x, s)}{t - s}$ for $s \neq t$. By (1) and the mean value theorem, $\forall t > s$ with $t, s \in \Delta_t$, $\exists u(s, t) \in [s, t] \ni \phi(s, t) = D_2F(x, u(s, t))$. Using this and (3) we have that for any $\varepsilon > 0$, $\forall b > 0$, $\exists \delta > 0$ and a nowhere-negative (in $\Delta_x$) $f(x)$ obeying $\int_b^\infty f(x)dx < \varepsilon$ such that if $|t - s| < \delta$ and $x \geq b$, then $|\phi(s, t) - D_2F(x, t)| = |D_2F(x, u(s, t)) - D_2F(x, t)| < f(x)$. From this and (2) it follows that for all $b > 0$, $\exists \delta > 0$, and a nowhere-negative (in $\Delta_x$) $f(x)$ obeying $\int_b^\infty f(x)dx < \varepsilon$ such that if $|t - s| < \delta$, then $\left|\int_b^\infty \phi(s, t)dx - \int_b^\infty D_2F(x, t)dx\right| \leq \int_b^\infty f(x)dx < \varepsilon$.

Taking the limit $s \to t$, noting that $\lim_{s \to t}\int_b^\infty \phi(s, t)dx = D_2\int_b^\infty F(x, t)dx$, and finally taking $\varepsilon \to 0$ with $b = \varepsilon$, we arrive at the desired result. QED.

The functions $F(x, t)$ of interest in this paper have the form $F(x, t) = x^t \log(x)^m e^{-cx}$ with $\mathrm{Re}(t) > -1$ and $c > 0$. For these functions it may be shown that the conditions of Thm. C.1 hold.

## D. EXPANDING ETA'S DOMAIN

To apply the T transform, the assumption that all $\eta_i < 0$ had to be made. Here we present a simple theorem that expands the region of validity of the various expressions derived in this paper to the region where any of the $\eta_i$ may be non-negative. We present the theorem for the single subset sum case only, although the multiple non-overlapping subset case and the contained overlap case may be handled in an almost identical manner.

Theorem D.1: If $\mathrm{Re}(\eta + \beta_1) > 0$, $\mathrm{Re}(\eta) \geq 0$ and $\mathrm{Re}(n_i) > -1$, $i = 1, \ldots, m$ then

$$I[\rho^\eta, \mathbf{n}] = \frac{\gamma_\mathbf{n}}{\Gamma(\beta_1)} \frac{\Gamma(\eta + \beta_1)}{\Gamma(\eta + \beta)}.$$

Proof: Note that $\eta > 0$ implies that there is an integer $q > 0$ and an $\bar{\eta} < 0$ such that $\eta = \bar{\eta} + q$. Thus $I[\rho^\eta, \mathbf{n}]$ may be rewritten as

$$I[\rho^\eta, \mathbf{n}] = \Sigma_{i \in \sigma} I[p_i \rho^{\bar{\eta} + q - 1}, \mathbf{n}] = \Sigma_{i \in \sigma} I[\rho^{\bar{\eta} + q - 1}, \mathbf{n} + \mathbf{e}_i],$$

where $[\mathbf{e}_i]_j = \delta_{ij}$ and $\delta$ is the Kronecker delta function. Iterate this operation $q$ times (removing one power from $\rho$ and summing with an increased count vector each time) to find

$$I[\rho^\eta, \mathbf{n}] = \Sigma_{i_1 \in \sigma} \ldots \Sigma_{i_q \in \sigma} I[\rho^{\bar{\eta}}, \mathbf{n} + \mathbf{e}_{i_1} + \ldots + \mathbf{e}_{i_q}].$$

Simplify this to yield

$$I[\rho^\eta, \mathbf{n}] = \Sigma_\mathbf{q} \binom{q}{\mathbf{q}} I[\rho^{\bar{\eta}}, \mathbf{n} + \mathbf{q}],$$

where the vector $\mathbf{q}$ has nonnegative integer components summing to $q$ with $q_i = 0$ for $i \notin \sigma$. The symbol $\binom{q}{\mathbf{q}} \equiv \frac{\Gamma(q+1)}{\Pi_{i=1}^{m}\Gamma(q_i+1)}$ is the usual multinomial coefficient. Since $\bar{\eta} < 0$, evaluate the integral $I[\rho^{\bar{\eta}}, \mathbf{n}+\mathbf{q}]$ using Thm. 12 with $k = 1$ (noting that $\beta_1$ and $\beta$ increase by q due to $\mathbf{q}$ being added to $\mathbf{n}$) to find $I[\rho^{\eta}, \mathbf{n}] = \frac{\gamma_{\mathbf{n}}\Gamma(\nu_1+\bar{\eta}+q)}{\Gamma(\nu_1)\Gamma(\nu+\bar{\eta}+q)}\Sigma_{\mathbf{q}}\binom{q}{\mathbf{q}}\gamma_{\mathbf{n}+\mathbf{q}}$(*). Now, we put $\Sigma_{\mathbf{q}}\binom{q}{\mathbf{q}}\gamma_{\mathbf{n}+\mathbf{q}}$ into closed form by noting that it is the discrete convolution product of the functions of $\Gamma(n_i+q_i+1)/\Gamma(q_i+1)$ of $q_i$ given by

$$\Sigma_{\mathbf{q}}\binom{q}{\mathbf{q}}\gamma_{\mathbf{n}+\mathbf{q}} = \Gamma(q+1)\,[\otimes_{i=1}^{m}\Gamma(n_i+q_i+1)/\Gamma(q_i+1)]\,(q).$$

Apply the Z transform convolution theorem (see App. B.2.) to find

$$\Sigma_{\mathbf{q}}\binom{q}{\mathbf{q}}\gamma_{\mathbf{n}} = \Gamma(q+1)Z^{-1}\,[\Pi_{i=1}^{m}Z\,[\Gamma(n_i+q_i+1)/\Gamma(q_i+1)]\,].$$

Note that $Z\left[\frac{\Gamma(n_i+q_i+1)}{\Gamma(q_i+1)}\right](z) = \Gamma(n_i+1)\,(1-z)^{-(n_i+1)}$ for $|z| < 1$ and substitute for the Z transforms to find $\Sigma_{\mathbf{q}}\binom{q}{\mathbf{q}}\gamma_{\mathbf{n}+\mathbf{q}} = \gamma_{\mathbf{n}}\frac{\Gamma(\beta_1+q)}{\Gamma(\beta_1)}$. Substituting this result in (*) and simplifying leads to the desired result. QED.

We resort to analytic continuation in the non-contained overlap case.

## E. EXISTENCE CONDITIONS

Here we present an example of a calculation for determining the conditions of existence of the various integrals $I[\cdot|\cdot]$ appearing in these papers. For the integrands of these papers, existence of these integrals depends upon the behavior of the singularities appearing at the edges of the region of integration.

Consider the single pair overlap intergral $I[\rho_1^{\eta_1}\rho_2^{\eta_2}, \mathbf{n}]$, where $\rho_1$, $\rho_2$, $\sigma_1$, and $\sigma_2$ are as in the definitions for Thm. 14a, with the minor change that $\sigma_{1+2}$ contains all $m$ indices, which may be made without loss of generality. We show that the conditions for existence of this integral are $\mathrm{Re}\,(\nu_i) > 0$, $i = 1, \ldots, m$, $\mathrm{Re}\,(\beta_1 + \eta_1) > 0$ and $\mathrm{Re}\,(\beta_2 + \eta_2) > 0$. Write the integral as

$$I\,[\rho_1^{\eta_1}\rho_2^{\eta_2}, \mathbf{n}] = \int d\mathbf{p}\ \Delta(\mathbf{p})\ \Theta(\mathbf{p})\ \rho_1^{\eta_1}\rho_2^{\eta_2}\ \Pi_{i=1}^{m} p_i^{n_i}. \tag{E.1}$$

The first condition, $\mathrm{Re}\,(\nu_i) > 0$, $i = 1, \ldots, m$, follows immediately from the fact that $\int dp_i\ p_i^{n_i}$ exists iff $\mathrm{Re}\,(\nu_i) > 0$ and the fact that any $p_i$ may independently be near zero for this particular overlap case. Now, either $\rho_1$ or $\rho_2$ may also be near zero. We consider the first case, in which $\rho_1$ is near zero, and use symmetry to supply the result for the second case. Letting $x \equiv \Sigma_{i \in \sigma_1} p_i$ and $y \equiv \Sigma_{i \in \sigma_{12}} p_i$, rewite Eq. (E.1) in a form that isolates $\rho_1$ as

$$I\,[\rho_1^{\eta_1}\rho_2^{\eta_2}, \mathbf{n}] = \int_0^1 dx\,x^{\eta_1}\int_0^x dy\,(1-x+y)^{\eta_2}\Big(\int d\mathbf{p}_{1-12}\,\delta(\Sigma_{1-12}p_i - (x-y))\,\Pi_{1-12}p_i^{n_i}$$

$$\times \int d\mathbf{p}_{12}\,\delta(\Sigma_{12}p_i - y)\,\Pi_{12}p_i^{n_i} \times \int d\mathbf{p}_{2-12}\,\delta(\Sigma_{2-12}p_i - (1-x+y))\,\Pi_{2-12}p_i^{n_i}\Big), \tag{E.2}$$

where here the subscript notation indicates the sets of indices involved, e.g. $1-12$ indicates $i \in \sigma_{1-12}$. Each of the three integrals over $\mathbf{p}$ in Eq. (E.2) may be done in closed form. Do these integrals using Thm. 9a and induction to find

$$I\,[\rho_1^{\eta_1}\rho_2^{\eta_2}, \mathbf{n}] = \frac{\gamma_{\mathbf{n}}}{\Gamma(\beta_{1-12})\Gamma(\beta_{12})\Gamma(\beta_{2-12})}$$

$$\times \int_0^1 dx\,x^{\eta_1}\int_0^x dy\,(x-y)^{\beta_{1-12}-1}\,y^{\beta_{12}-1}\,(1-x+y)^{\eta_2+\beta_{2-12}-1}. \tag{E.3}$$

Apply the binomial theorem in (E.3) to expand two of the three factors in the integrand, $(x-y)^{\beta_{1-12}-1}$ and $(1-x+y)^{\eta_2+\beta_{2-12}-1}$, in series. Using these series, note that each term in the series for $(x-y)^{\beta_{1-12}-1}$ will contribute the same power of $x$ after the integration over $y$, while the terms of the series for $(1-x+y)^{\eta_2+\beta_{2-12}-1}$ contribute increasing powers of x after the integration over $y$. Note also that if the lowest-power-of-$x$ term is integrable over $x$ in a region containing $0$, then all terms are. Thus, the worst case occurs with the constant term from the binomial series for $(1-x+y)^{\eta_2+\beta_{2-12}-1}$. After integration over $y$ with this constant term, and considering the small $x$ region of integration, we are left with the integral over $x$ given by

$$C\int_0^{\underline{x}} dx\ x^{\eta_1+\beta_1-1}, \tag{E.4}$$

where $0<\underline{x}<1$, and $C$ is a constant. This integral exists for $\mathrm{Re}(\beta_1+\eta_1)>0$. This, symmetry, and the first condition (given by $\mathrm{Re}(\nu_1)>0$, $i = 1, \ldots, m$) establish the result. The method for more complicated overlap structures is also indicated by this discussion.

The discussion above is of interest in another way: it provides a general method for finding multiple overlap integrals without the use of transform theory.

## F. DERIVATIVES OF OVERLAP CONVOLUTIONS - POLES

In this appendix we find derivatives with respect to $\eta$ of expressions such as $\Gamma(k-\eta)/\Gamma(-\eta)$, where $k \in \{0, 1, \ldots\}$, and $\eta$ may be any number within the constraints of existence. We consider the various cases that arise when some combination of poles occurs and demonstrate the various simplified expressions for the derivatives in these cases.

When $\eta$ is not an integer, there are no poles in either $\Gamma(k-\eta)$ or $\Gamma(-\eta)$ and the usual derivative expressions hold. When $\eta$ is an integer and $\eta<0$, the usual expressions also hold since $k\ge 0$ and therefore $k-\eta>0$. The case where the usual expressions hold will be denoted as case 0.

When $\eta$ is an integer and $\eta\ge 0$, there are two cases are of interest. The first, case 1, occurs when $\eta\ge 0$ and $k-\eta>0$, so that there are poles in the denominator of $\Gamma(k-\eta)/\Gamma(-\eta)$ only. The second, case 2 occurs when $\eta \ge 0$ and $k - \eta \le 0$, so that there are poles in both the numerator and the denominator.

In order to find expressions for the derivatives in cases 1 and 2 we use the following facts. 1) The only singularities of the gamma function $\Gamma(x)$ are simple poles at $x = -n, n = 0, 1, \ldots$ with residues $(-1)^n / n!$ respectively. 2) $\Delta\Phi^{(n)}(k - \eta, -\eta) = (-1)^{n-1}\Gamma(n)\Sigma_{i=1}^{k} \frac{1}{(k - i - \eta)^n}$ whenever the expressions exist. (The identity $\Gamma(k - \eta) / \Gamma(-\eta) = \Pi_{i=1}^{k} (k - i - \eta)$ may be used in deriving this.) 3) $\Gamma(k - \eta) / \Gamma(-\eta)$ is the representation (away from the poles in the gamma functions) of an everywhere-analytic function (note that $k \ge 0$ is still assumed). Using these facts, the expressions for cases 1 and 2 are found by substituting $\xi = \eta + \varepsilon$ for $\eta$ (now restricted by the conditions of cases 1 and 2 to be a nonnegative integer) in the corresponding case 0 expressions and taking the $\varepsilon = 0$ limit.

Case 0: $\eta$ non-integer, or $\eta$ an integer with $\eta < 0$ and $k - \eta > 0$. There are no poles in the numerator or denominator. The first derivative is given by

$$\partial_\eta^1 \left[ \frac{\Gamma(k - \eta)}{\Gamma(-\eta)} \right] = -\frac{\Gamma(k - \eta)}{\Gamma(-\eta)} \Delta\Phi^{(1)}(k - \eta, -\eta). \tag{F.1}$$

The $r^{th}$ derivative may be found by iteration, using Eq. (F.1) and the recursion relation

$$\partial_\eta^1 \Phi^{(n)}(k - \eta) = -\Phi^{(n+1)}(k - \eta). \tag{F.2}$$

For example, taking the derivative of Eq. (F.1), applying Eq. (F.2) with Eq. (F.1) in the process, yields the second derivative

$$\partial_\eta^2 \left[ \frac{\Gamma(k - \eta)}{\Gamma(-\eta)} \right] = \frac{\Gamma(k - \eta)}{\Gamma(-\eta)} [\Delta\Phi^{(1)}(k - \eta, -\eta)^2 + \Delta\Phi^{(2)}(k - \eta, -\eta)] . \tag{F.3}$$

Case 1: $\eta$ an integer, $\eta \ge 0$ and $k - \eta > 0$. The denominator contains a pole.

The zeroth derivative is 0. Taking the appropriate limit in Eq. (F.1) gives us the first

derivative

$$(-1)^{\eta+1}\eta!\Gamma(k-\eta), \tag{F.4}$$

while taking the appropriate limit in Eq. (F.3) yields the second derivative

$$(-1)^{\eta}2\eta!\Gamma(k-\eta)\Sigma^{k}_{i=1,\, i\neq k-\eta}(k-i-\eta)^{-1}. \tag{F.5}$$

Case 2: $\eta$ integer, $\eta \geq 0$ and $k-\eta \leq 0$: Both the numerator and denominator contain poles. The zeroth derivative is simply $\frac{(-1)^{k}\eta!}{(\eta-k)!}$. Taking the appropriate limit in Eq. (F.1) gives us the first derivative

$$(-1)^{k+1}\frac{\eta!}{(\eta-k)!}\Delta\Phi^{(1)}(k-\eta, -\eta), \tag{F.6}$$

while taking the appropriate limit in Eq. (F.3) yields the second derivative

$$(-1)^{k}\frac{\eta!}{(\eta-k)!}[\Delta\Phi^{(1)}(k-\eta, -\eta)^{2}+\Delta\Phi^{(2)}(k-\eta, -\eta)]. \tag{F.7}$$

## G. MULTIPLE OVERLAP RESULTS

Here we define the general *Convolution Form* (CF) notation and demonstrate the results stated without proof in Sec. 2e. Let $\mathbf{k} = (k_1, \ldots, k_n)$ be a vector of nonnegative summation indices, (with n the number of such indices) and let $C(\cdot)$ be any function of an n-vector of nonnegative integers. Further, let $t^{\mathbf{k}} \equiv \Pi^{n}_{i=1}t_i^{k_i}$, $t_{\mathbf{a}} \equiv \Sigma^{n}_{i=1}a_i t_i$, where $\mathbf{a} = (a_1, \ldots, a_n)$ with components in the reals, and let $\alpha_{\mathbf{k}}$ be a complex number indexed by $\mathbf{k}$.

Define the CF symbol $(\mathbf{a};\mathbf{b})$ to be the set of expressions of the form

$$\exp(-\tau t_{\mathbf{a}})\,\Sigma_{\mathbf{k}}C(\mathbf{k})\tau^{\alpha_{\mathbf{k}}-1}t^{\mathbf{k}}, \tag{G.1}$$

where $\tau$ is known as the *convolution variable* and where the *occurrence* indicator vector $\mathbf{b}$ has components in $\{0, 1\}$ and $b_i = 1$ if and only if $t_i$ *occurs* in the summation $\Sigma_{\mathbf{k}} C(\mathbf{k}) \tau^{\alpha_{\mathbf{k}} - 1} t^{\mathbf{k}}$, i.e. $b_i = 1$ iff $\exists \mathbf{k} \ni C(\mathbf{k}) \neq 0$ and $k_i > 0$. All the aspects of $C(\mathbf{k})$ that are relevant to the analysis of Sec. 2e are expressed by the vector $\mathbf{b}$. A particular member of the CF $(\mathbf{a};\mathbf{b})$ will be written as $(\mathbf{a};\mathbf{b})_u$, i.e. u indexes the CF $(\mathbf{a};\mathbf{b})$.

In order to represent convolutions of CF expressions, let $[(\mathbf{a};\mathbf{b}), (\mathbf{c};\mathbf{d})]$ be the set of all expressions that result from the convolution of two CF members from $(\mathbf{a};\mathbf{b})$ and $(\mathbf{c};\mathbf{d})$ respectively, i.e.

$$[(\mathbf{a};\mathbf{b}), (\mathbf{c};\mathbf{d})] \equiv \{\int_0^\tau (\mathbf{a};\mathbf{b})_u(x)\ (\mathbf{c};\mathbf{d})_w(\tau - x)\ dx\}, \tag{G.2}$$

where by $(\mathbf{a};\mathbf{b})_u(x)$ we mean $(\mathbf{a};\mathbf{b})_u$ with its convolution variable evaluated at $x$, and where $u$ and $w$ range over all indices for the CF's $(\mathbf{a};\mathbf{b})$ and $(\mathbf{c};\mathbf{d})$ respectively.

Now we show that the result of the convolution of any two members of CF's may be written as a member of some CF. More precisely, we prove Eq. (13) of Sec. 2e,

$$[(\mathbf{a};\mathbf{b}), (\mathbf{c};\mathbf{d})] \subset (\mathbf{c}\ ;\ \mathbf{b} \vee \mathbf{d} \vee \mathrm{nz}(\mathbf{c} - \mathbf{a})). \tag{G.3}$$

To prove Eq. (G.3) start by noting that each integrated convolution in Eq. (G.2) has the representation (see App. A, Eq. (A.2))

$$e^{-\tau t_{\mathbf{c}}} \Sigma_{\mathbf{j}} \Sigma_{\mathbf{k}} C(\mathbf{j}, \mathbf{k}) t^{\mathbf{j}} t^{\mathbf{k}} \tau^{\alpha_{\mathbf{j}} + \alpha_{\mathbf{k}} - 1} \frac{\Gamma(\alpha_{\mathbf{j}})\Gamma(\alpha_{\mathbf{k}})}{\Gamma(\alpha_{\mathbf{j}} + \alpha_{\mathbf{k}})} \times {}_1F_1(\alpha_{\mathbf{j}}; \alpha_{\mathbf{j}} + \alpha_{\mathbf{k}}; \tau t_{\mathbf{c} - \mathbf{a}}). \tag{G.4}$$

In Eq. (G.4), as before, $\mathbf{j}$ and $\mathbf{k}$ are vectors of nonnegative summation indices, $C(\cdot, \cdot)$ is a function of the vectors $\mathbf{j}$ and $\mathbf{k}$ (specifically, it is the product of the C's appearing in the two CF's $(\mathbf{a};\mathbf{b})_u(x)$ and $(\mathbf{c};\mathbf{d})_w(x)$), $t^{\mathbf{j}} \equiv \Pi_{i=1}^n t_i^{[\mathbf{j}]_i}$ (and similarly for $t^{\mathbf{k}}$), $t_{\mathbf{c}} \equiv \Sigma_{i=1}^n c_i t_i$, and $t_{\mathbf{c} - \mathbf{a}} \equiv \Sigma_{i=1}^n (c_i - a_i) t_i$. Now, expand the hypergeometric ${}_1F_1$ in the series (see App. A)

$${}_1F_1(\alpha_{\mathbf{j}}; \alpha_{\mathbf{j}} + \alpha_{\mathbf{k}}; \tau t_{\mathbf{c} - \mathbf{a}}) = \frac{\Gamma(\alpha_{\mathbf{j}} + \alpha_{\mathbf{k}})}{\Gamma(\alpha_{\mathbf{j}})\Gamma(\alpha_{\mathbf{k}})} \Sigma_{i=0}^{\infty} \frac{(\alpha_{\mathbf{j}})_i}{(\alpha_{\mathbf{j}} + \alpha_{\mathbf{k}})_i} \frac{(\tau t_{\mathbf{c} - \mathbf{a}})^i}{i!}. \tag{G.5}$$

Substitute Eq. (G.5) into Eq. (G.4) and note the following four things: (1) if $b_i = 1$ then $t_i$ occurs in Eq. (G.4), (2) if $d_i = 1$ then $t_i$ occurs in Eq. (G.4), (3) if $c_i - a_i \neq 0$ then $t_i$ occurs in Eq. (G.4), and (4) if none of the cases 1-3 hold then $t_i$ does not occur in Eq. (G.4). Thus, the occurrence indicator vector for the result of the convolution Eq. (G.4) is $(\mathbf{c}\ ;\ \mathbf{b} \vee \mathbf{d} \vee \mathrm{nz}(\mathbf{c} - \mathbf{a}))$ . Noting also that the constant in the exponential's argument is $t_{\mathbf{c}}$ establishes Eq. (G.3).

Finally, Eq. (14) of Sec. 2e follows immediately from the definition of the CF in Eq. (G.1) and the properties of the T transform in App. A.

**FOOTNOTES**

[1] An alternate proof of Thm. 12 leads to an identity. Where the inverse T transform is applied in the proof of Thm. 12, instead find the convolution $(\tau^{\beta_1 - 1} e^{-\tau t} \otimes \tau^{\bar{\beta} - 1})$ using Thm. 10.2 and express it in terms of ${}_1F_1$. Now, do the inverse transform and equate this result to the result of Thm. 12 to find *Gauss's identity*: ${}_2F_1((a, b); c; 1) = \dfrac{\Gamma(c)\Gamma(c-a-b)}{\Gamma(c-a)\Gamma(c-b)}$.

[2] An alternate proof of Thm. 14a leads to another identity. Instead of applying Thm. 11, apply Thm. 10.1 to the first two terms of the pairwise overlap convolution $p^{\beta_{1-12} - 1} e^{-pt_1} \otimes p^{\beta_{12} - 1} e^{-p(t_1 + t_2)}$ and immediately take the inverse $T_1$ transform. Now, do the final convolution with $p^{\beta_{2-12} - 1} e^{-pt_2}$ and take the inverse $T_2$ transform. Note that there is only a single summation in the result, whereas in the result in Thm. 14a there are two summations. On the other hand, the convolution $p^{\beta_{12} - 1} e^{-p(t_1 + t_2)} \otimes p^{\beta_{2-12} - 1} e^{-pt_2}$ can be taken first, followed by a convolution with $p^{\beta_{1-12} - 1} e^{-pt_1}$, effectively interchanging indices 1 and 2. Equating these two single-sum forms gives the identity

$$
{}_3F_2[a_1, a_2, a_3; b_1, b_2; 1] = \frac{\Gamma(b_1)\Gamma(b_2)\Gamma(b_1 + b_2 - (a_1 + a_2 + a_3))}{\Gamma(a_2)\Gamma(b_1 + b_2 - (a_1 + a_2))\Gamma(b_1 + b_2 - (a_2 + a_3))}
$$

$$
\times {}_3F_2[b_1 - a_2, b_1 + b_2 - (a_1 + a_2 + a_3), b_2 - a_2; b_1 + b_2 - (a_1 + a_2), b_1 + b_2 - (a_2 + a_3); 1],
$$

while equating either of the single-sum results just described to the original result of Thm. 14a yields Gauss's identity, discussed in footnote 1.

[3] Utilizing Gauss's identity (see footnote 1 and [9], Eq. 15.1.1) provides further simplification in Thm. 15a for cases 15a.2 and 15a.4. These simplifications are due to simplifications appearing in $F^{(10)}$ and $F^{(01)}$ respectively. The choice of the form of the results presented was made considering the simplicity of the results and consistency between the results.